\documentclass[12pt]{iopart}
\usepackage{graphicx} 

\begin{document} 
%\jl{20} 
 
\title{On Tamm's problem in the Vavilov-Cherenkov radiation theory}

\author{G N Afanasiev\dag
V G Kartavenko\dag\
and Yu P Stepanovsky\ddag}

\address{\dag\ Bogoliubov Laboratory of Theoretical Physics, 
Joint Institute for Nuclear Research,  
Dubna, 141980, Moscow District, Russia} 
\address{\ddag\ The Institute of Physics and Technology,
Kharkov, Ukraine} 

\begin{abstract}
We analyse the well-known Tamm problem treating  the charge motion
on a finite space interval with the velocity exceeding light velocity
in medium. By comparing Tamm's formulae with the exact ones we prove that
former do not properly describe Cherenkov radiation terms. 
We also investigate Tamm's formula $\cos\theta=1/\beta n$ defining 
the position  of maximum of the field
strengths  Fourier components for the infinite uniform motion of a charge.
Numerical analysis of the Fourier components of field strengths shows
that they have a pronounced maximum  at $\cos\theta=1/\beta n$ 
only for the charge
motion on the infinitely small interval. As the latter grows, many maxima
appear. For the  charge motion on an infinite  interval there is infinite
number of maxima of the same amplitude. The quantum analysis of Tamm's formula
leads to the same results.
\end{abstract}

% 
%  Uncomment out if preprint format required 
% 
\maketitle 

\section{Introduction} 

In 1888 O. Heaviside considered an infinite charge motion in the nondispersive
dielectric infinite medium [1]. He showed that a specific radiation arises
when the charge velocity  $v$ exceeds the light velocity in medium $c_n$.
This radiation is confined to the cone with a vertex angle 
$\sin\theta_s=1/\beta_n$. Here $\beta_n=v/c_n$. 
The Poynting vector being perpendicular to this cone has the angle
$$\cos\theta_c=1/\beta_n\eqno(1.1)$$ with the motion axis. This radiation was
experimentally observed by P.A. Cherenkov in 1934 [2]. 
Unfortunately, Heaviside's studies had been forgotten until 1974 when 
they were revived by A.A. Tyapkin [3] and T.R. Kaiser [4].

I.E. Tamm and I.M. Frank [5] without knowing the previous Heaviside 
investigations explained Cherenkov's experiments  solving the Maxwell 
equations in the Fourier representation and subsequently returning to 
the usual space-time representation.
The use of the Fourier representation permitted them to treat 
the dispersive media as well. 
For the non-dispersive media they confirmed the validity of Eq.(1.1)
defining the direction of the Cherenkov radiation.

In 1939 I.E.Tamm [6] considered the uniform motion of a point charge on the
finite space interval with the velocity $v$ exceeding the light velocity
in medium $c_n$. Here $c_n=c/n(\omega)$, $n(\omega)$ 
is the frequency-dependent
refraction index of the medium. He showed that Fourier components of
electromagnetic field strengths have a sharp maximum at the angle
$$\cos\theta_T=1/\beta_n \eqno(1.2)$$ with the motion axis.
Here $\beta_n=v/c_n(\omega)$. Later (see, e.g., [7]) Eq.(1.2) has been
extended to the  charge motion in an infinite medium.

On the other hand, in Ref. [8] the uniform motion of a point
charge was considered in an infinite dispersive medium with 
a one-pole electric penetrability
chosen in a standard way [9]:
$$\epsilon(\omega)=1+\frac{\omega_L^2}{\omega_0^2-\omega^2}.\eqno(1.3)$$
This expression  is a suitable extrapolation between the static case
$\epsilon(0)=1+\omega_L^2/\omega_0^2$ and the
high-frequency limit $\epsilon(\infty)=1$.
The electromagnetic potentials, field strengths and the energy flux were
evaluated on the surface of a cylinder co-axial 
with the charge axis motion $z$.
They had the main maximum at those points of the cylinder surface 
where in the absence
of dispersion it is intersected by the Cherenkov singular cone and
smaller maxima in the interior of this cone. On the other hand,
the Fourier transforms of these quantities were oscillating functions of $z$
and, therefore, of the scattering angle $\theta$ ($z=r\cos\theta$) 
without a pronounced maximum at $\cos\theta=1/\beta_n$. 
This disagrees with the validity of Eq.(1.2)
(not (1.1)) for the infinite charge motion.

Lawson [10,11] qualitatively analyzing Tamm's formula concluded that 
the distinction between the Cherenkov radiation and bremsstrahlung 
completely disappears
for the small motion interval and is maximal for the large one.

Further, Zrelov and Ruzicka ([12,13]) numerically investigating Tamm's problem
came to the paradoxical result that Tamm's formulae (which, as they
believed, describe the Cherenkov radiation) can be interpreted 
as the interference of two bremsstrahlung ($BS$) waves emitted 
at the beginning and end of motion.

Slightly later, the exact solution of the same problem in the absence of
dispersion has been found in [14]. 
It was shown there that Cherenkov's radiation
exists for any motion interval and by no means can be reduced 
to the interference  of two $BS$ waves. 
This is also confirmed by the results of Ref. [15]  where 
the exact electromagnetic field of a point charge moving with a constant 
acceleration in medium has been found (the motion
begins either from the state of rest or terminates  with it).

These inconsistencies and the fact that formula (1.2) for the Fourier
components  is widely used for
the identification of the Cherenkov radiation even for the uniform
charge motion in an infinite medium enable us to reexamine Tamm's problem anew.

The plan of our exposition is as follows. In Sect. 2, we reproduce step by
step the derivation of Tamm's formulae. 
In Sect. 3, by comparing them  with exact 
ones we prove that Tamm's approximate formulae  do not describe  Cherenkov's
radiation properly. The reason for this is due to the approximations 
involved in their
derivation. In Sect. 4, we analyze the validity of Tamm's formula (1.2) for
different intervals of charge motion. 
We conclude that it is certainly valid for
small intervals and breaks for larger ones. This is also supported by the
numerical calculations and analytical formula available 
for the infinite charge motion. On the other
hand, the Tamm-Frank formula (1.1) is valid even for the dispersive media:
it approximately defines the position of main intensity maximum in the
usual space-time representation ([8]). Quantum analysis of Tamm's formula
given in Sect. 5 definitely supports the results of previous sections.
A short discussion of the results obtained is given in section 6.

Some precaution is needed. When experimentally investigating a charge motion
on a finite interval [16], one usually considers an electron beam entering a
thin transparent slab from vacuum, 
its propagation inside the slab and the subsequent passing
into the vacuum on the other side of the slab. 
The so-called transition radiation
[17] arises on the slab interfaces. 
In this investigation we deal with a pure
Tamm's problem: electron starts at a given point in  medium, propagates with
a given velocity and then stops at a second point. This may be realized, e.g.,
for the electron propagation in water where the distance  between successive
scatters is $\approx 1\mu m$, which is approximately twice the wavelength of
the visible Cherenkov radiation [18]. Another realization of Tamm' problem
is a $\beta$ decay followed by the nuclear capture [7,13].

\section{Tamm's problem}
Tamm considered the following problem. The point charge rests at the point
$z=-z_0$ of the $z$ axis up to a moment $t=-t_0$. In the time interval
$-t_0<t<t_0$ it uniformly moves along the $z$ axis with the velocity $v$
greater than the light velocity in medium $c_n$.
For $t>t_0$ the charge again rests at the point $z=z_0$. The non-vanishing
$z$ Fourier component of the vector potential (VP) is given by
$$A_\omega=\frac{1}{c}\int_{-z_0}^{z_0}\frac{1}{R}
j_\omega(x',y',z')\exp{(-i\omega R/c)} dx'dy'dz',$$
where $R=[(x-x')^2+(y-y')^2+(z-z')^2]^{1/2}$, $j_\omega=0$ for $z'<-z_0$ and
$z'>z_0$ and $j_\omega=e\delta(x')\delta(y')\exp{(-i\omega z'/v)}/2\pi$ 
for $-z_0<z'<z_0$.
Inserting all this into $A_\omega$ and integrating over $x'$ and $y'$ one gets
$$A_\omega(x,y,z)=\frac{e}{2\pi c}\int _{-z_0}^{z_0}\frac{d z'}{R}
\exp{[-i\omega(\frac{z'}{v}+\frac{R}{c_n})]},$$
$$ R=[\rho^2+(z-z')^2]^{1/2},\quad \rho^2=x^2+y^2.\eqno(2.1)$$
At large distances from the charge ($R>>z_0$) one has: 
$R=R_0-z'\cos\theta,\; \cos\theta=z/R_0$.
Inserting this into (2.1) and integrating over $z'$ one gets
$$A_\omega(\rho,z)=\frac{e\beta q(\omega)}{\pi R_0\omega} 
\exp{(-i\omega R_0/c_n)},\quad q(\omega)=\frac{\sin{[\omega t_0
(1-\beta_n\cos\theta)]}}{1-\beta_n\cos\theta}.\eqno(2.2)$$
Now we evaluate the field strengths. 
In the wave zone where $R_0>>c/n\omega$ one obtains
$$H_\phi=-\frac{2e\beta}{\pi c R_0}\sin\theta\int_{0}^{\infty}nq(\omega)
\sin[\omega(t-R_0/c_n)] d\omega,$$
$$E_\rho=-\frac{2e\beta}{\pi c R_0}\sin\theta\cos\theta\int_{0}^{\infty}
q(\omega)\sin[\omega(t-R_0/c_n)] d\omega,$$
$$E_z=\frac{2e\beta}{\pi c R_0}\sin^2\theta\int_{0}^{\infty}
q(\omega)\sin[\omega(t-R_0/c_n)]d\omega.\eqno(2.3)$$
It should be noted that  only the $\theta$ spherical component of $\vec E$
differs from zero
$$E_r=0,\quad E_\theta=-\frac{2e\beta}{\pi c R_0}\sin\theta\int_{0}^{\infty}
q(\omega)\sin[\omega(t-R_0/c_n)] d\omega.$$
Consider now the function $q(\omega)$. For $\omega t_0>>1$ 
it goes into $\pi\delta(1-\beta_n\cos\theta)$.
This means that under these conditions $\vec E_\omega$ and 
$\vec H_\omega$ have a sharp
maximum for $1-\beta_n\cos\theta=0$. 
Or, in other words, photons with the energy
$\hbar\omega $ should be observed at the angle $\cos\theta=1/\beta_n$.

The energy flux through the sphere of the radius $R_0$ is
$$W=R_0^2\int S_r d\omega,\quad S_r=\frac{c}{4\pi}E_\theta H\phi.$$
Inserting  $E_\theta$ and $H_\phi$ one obtains
$$W=\frac{2e^2\beta^2}{\pi c}\int_{0}^{\infty}nJ(\omega) d\omega,
\quad J(\omega)=\int_{0}^{\infty} q^2\sin\theta d\theta.$$
For $\omega t_0>>1$, $J$ can be evaluated in a closed form
$$J=J_{BS}=\frac{1}{\beta^2 n^2}(\ln\frac{1+\beta_n}{|1-\beta_n|}-2\beta_n)
\quad {\rm for}\quad \beta_n<1\quad {\rm and}$$
$$J=J_{BS}+J_{Ch},\quad J_{Ch}=\frac{\pi\omega t_0}{\beta_n}(1-
\frac{1}{\beta_n^2})\quad {\rm for}\quad \beta_n>1.\eqno(2.4)$$
Tamm identified $J_{BS}$ with the spectral distribution of the bremsstrahlung
$BS$ , arising from instant acceleration and deceleration of
the charge at the moments $\pm t_0$, resp. On the other hand, $J_{Ch}$ was
identified with the spectral distribution of the Cherenkov radiation. This is
supported by the fact that
$$W_{Ch}=\frac{2e^2\beta^2}{\pi c}\int_{0}^{\infty}nJ_{Ch}(\omega) 
d\omega=\frac{2e^2\beta^2 t_0}{ c}\int_{\beta_n>1}\omega 
d\omega (1-\frac{1}{\beta_n^2}).\eqno(2.5)$$
strongly resembles the famous Frank-Tamm formula [5] for an infinite medium
obtained in a quite different way.

In the absence of dispersion Eqs.(2.3) are easily integrated:
$$H_\phi^T=-\frac{e\beta\sin\theta}{R_0(1-\beta_n\cos\theta)}
\{\delta[c_n(t-t_0)-R_0+z_0\cos\theta]-\delta[c_n(t+t_0)
-R_0-z_0\cos\theta] \},$$
$$E_\theta^T=-\frac{e\beta\sin\theta}{R_0n(1-\beta_n\cos\theta)}
\{\delta[c_n(t-t_0)-R_0+z_0\cos\theta]-\delta[c_n(t+t_0)-R_0-z_0
\cos\theta] \}\eqno(2.6).$$
Superscript $T$ means that these expressions originate from  Tamm's field
strengths (2.2).

\section{Comparison with exact solution}
\subsection{Exact solution}
On the other hand, in Ref. [14] there was given an exact solution 
of the treated problem 
(i.e., the superluminal charge motion on the finite space interval)
in the absence of dispersion.
It is assumed that a point charge moves on the interval $(-z_0,z_0)$ 
lying inside $S_0$.
The charge  motion begins at the  moment $t=-t_0=-z_0/v$ 
and terminates at the moment $t=t_0=z_0/v$.
For convenience we shall refer  to the $BS$ shock waves emitted at the
beginning of the charge motion ($t=-t_0$) and at its termination ($t=t_0$)
as to the $BS_1$ and $BS_2$ shock waves, resp.

In the wave zone the field strengths are of the form ([14])
$$\vec E=\vec E_{BS}+\vec E_{Ch},\quad \vec E_{BS}=\vec E_{BS}^{(1)}+
\vec E_{BS}^{(2)} \quad \vec H=\vec H_{BS}+\vec H_{Ch}, $$
$$\vec H=H_\phi\vec n_\phi,\quad H_\phi=H_{BS}+H_{Ch},
\quad H_{BS}= H_{BS}^{(1)}+H_{BS}^{(2)}.\eqno(3.1) $$
Here
$$\vec E_{BS}^{(1)}=-\frac{e\beta}{n}\frac{\delta[c_n(t+t_0)-r_1]}
{\beta_n(z+z_0)-r_1}\frac{r\sin\theta}{r_1}\vec n_{\theta}^{(1)},
\quad \vec E_{BS}^{(2)}=\frac{e\beta}{n}\frac{\delta[c_n(t-t_0)-r_2]}
{\beta_n(z-z_0)-r_2}\frac{r\sin\theta}{r_2}\vec n_{\theta}^{(2)},$$
$$\vec E_{Ch}=\frac{2}{\epsilon r_m\gamma_n}\delta(c_n t-R_m)
\Theta(\rho\gamma_n+z_0-z)\Theta(-\rho\gamma_n+z_0+z) \vec n_m,$$
$$H_{BS}^{(1)}=-e\beta\frac{\delta[c_n(t+t_0)-r_1]}{\beta_n(z+z_0)-r_1}
\frac{r\sin\theta}{r_1},\quad H_{BS}^{(2)}=e\beta
\frac{\delta[c_n(t-t_0)-r_2]}{\beta_n(z-z_0)-r_2}\frac{r\sin\theta}{r_2},$$
$$H_{Ch}=\frac{2}{ r_m\gamma_n\sqrt{\epsilon\mu}}\delta(c_n t-R_m)
\Theta(\rho\gamma_n+z_0-z)\Theta(-\rho\gamma_n+z_0+z) \vec n_\phi,$$
$$\gamma_n=|1-\beta_n^2|^{-1/2},\quad r_1=[(z+z_0)^2+\rho^2]^{1/2},
\quad r_2=[(z-z_0)^2+\rho^2]^{1/2},$$
$$ r_m=[(z-vt)^2-\rho^2/\gamma_n^2]^{1/2},\quad R_m=(z+\rho/
\gamma_n)/\beta_n,$$
$$n_\theta^{(1)}=[\vec n_\rho(z+z_0)-\rho\vec n_z]/r_1,
\quad n_\theta^{(2)}=[\vec n_\rho(z-z_0)-\rho\vec n_z]/r_2,
\quad \vec n_m=(\vec n_\rho-n_z/\gamma_n )/\beta_n.$$
The meaning of this notation is as follows:
$\Theta(x)$ is a step function ($\Theta(x)=0$ for $x<0$ and 
$\Theta(x)=1$ for $x>0$;
$r=\sqrt{z^2+\rho^2}$ is the distance of the observation point from the
origin (it coincides with Tamm's $R_0$);
$r_1=\sqrt{(z+z_0)^2+\rho^2}$ and $r_2=\sqrt{(z-z_0)^2+\rho^2} $ are the
distances of the observation point from the points
of the motion axis where the instant acceleration 
(at $t=-t_0$) and deceleration (at $t=t_0$) take place.
Correspondingly, $\delta$ functions $\delta[c_n(t+t_0)-r_1]$ and 
$\delta[c_n(t-t_0)-r_2]$
describe spherical $BS$ shock waves emitted at these moments;
$n_\theta^{(1)}$ and $n_\theta^{(1)}$ are the unit vectors 
tangent to the above
spherical waves and lying in the $\phi=const$ plane;
$\vec E_{BS}^{(1)}$, $\vec E_{BS}^{(2)}$, $\vec H_{BS}^{(1)}$ and 
$\vec H_{BS}^{(2)}$
are the electric and magnetic field strengths of the BS shock waves.
The function $\delta(c_n t-R_m)$ describes the position  of the Cherenkov
shock  wave ($CSW$). The inequalities
$R_m<c_nt$ and $R_m>c_nt$  correspond to the points lying inside the VC cone
and outside it, resp.;
$\vec n_m$ is the vector lying on the surface of 
the Vavilov-Cherenkov (VC) cone; $r_m$ is the
so-called Cherenkov singularity: $r_m=0$ on the VC cone surface;
$\vec E_{Ch}$ and $\vec H_{Ch}$ are the electric and magnetic field strengths
describing  $CSW$; $\vec E_{Ch}$ and $\vec H_{Ch}$ are infinite
on the surface of the VC cone and vanish outside it. Inside the VC cone
$\vec E_{Ch}$ and $\vec H_{Ch}$ decrease as $r^{-2}$
at large distances and, therefore, do not give contribution in
the wave zone where only the radiation terms are essential.
\subsection{Comparison with Tamm's solution}

At large distances one may develop $r_1$ and $r_2$ in (3.1):
$r_1=r+z_0\cos\theta,\quad r_2=r-z_0\cos\theta$. 
Here $r=R_0=[\rho^2+z^2]^{1/2}$.
Neglecting  $z_0$ compared with $r$ in the denominators of
$\vec E_{BS}$ and $\vec H_{BS}$ in (3.1), one gets
$$ \vec E_T=\vec E_{BS},\quad \vec H_T=\vec H_{BS},\quad
\vec E=\vec E_T+\vec E_{Ch},\quad \vec H=\vec H_T+\vec H_{Ch},  $$
where $\vec E_T$ and $\vec H_T$ are the same as in Eq.(2.6).
This means that Tamm's field strengths (2.6) describe only the bremsstrahlung
and do not contain the Cherenkov singular terms. Correspondingly, the maxima
of their Fourier transforms refer to the BS radiation.\\
To elucidate why the Cherenkov radiation is absent in Eqs. (2.3), we
consider the  product of two $\Theta$ functions entering into 
the definition (3.1)
of Cherenkov field strengths $\vec E_{Ch}$ and $\vec H_{Ch}$:
$$\Theta(\rho\gamma_n+z_0-z)\Theta(-\rho\gamma_n+z_0+z).$$
If for
$$z_0<< \rho\gamma_n-z=r(\gamma_n\sin\theta-\cos\theta)\eqno(3.2)$$
one naively neglects the term $z_0$ inside the $\Theta$ functions, 
the product of two $\Theta$ functions reduces to 
$\Theta(\rho\gamma_n-z)\Theta(-\rho\gamma_n+z)$ that is equal to zero.
In this case the Cherenkov radiation drops out.

We prove now that essentially the same approximation was implicitly made 
during the transition from (2.1) to (2.2). 
When changing $R$ under the sign of exponent  in (2.1) by 
$R_0-z'\cos\theta$ it was implicitly assumed that the quadratic term 
in the development of $R$ is small
as compared to the linear one. 
Consider this more carefully. We develop $R$
up to the second order:
$$R\approx R_0-z'\cos\theta+\frac{z'^2}{2R}\sin^2\theta.$$
Under the sign of exponent in (2.1) the following terms appear
$$\frac{z'}{v}+\frac{1}{c_n}(R_0-z'\cos\theta + 
\frac{z'^2}{2R_0}\sin^2\theta).$$
We collect terms involving $z'$
$$\frac{z'}{c_n} [( \frac{1}{\beta_n}-\cos\theta) + 
\frac{z'}{2R_0}\sin^2\theta].$$
Taking for $z'$ its maximal value $z_0$, we present the condition
for  the second term in the development of $R$ to be small  in the form
$$z_0<<2R_0(\frac{1}{\beta_n}-\cos\theta)/\sin^2\theta$$
It is seen that the right-hand side of this equation and that of Eq.(3.2)
vanish for
$ \cos\theta=1/\beta_n $, i.e., at the angle where 
the Cherenkov radiation exists.
This means that the absence of the Cherenkov radiation in Eqs. (2.3) 
is due to  the omission of second-order terms in the development of $R$ 
under the exponent in (2.1).

\subsection{Space distribution of shock waves}
Consider space distribution of the electromagnetic field (EMF) 
at the fixed moment of time. It is convenient to deal with the space 
distribution of the magnetic vector potential
rather than with that of field strengths which are the space-time derivatives
of electromagnetic potentials. \\
The exact electromagnetic potentials are equal to ([14])
$$\Phi= \Phi_1+\Phi_2+\Phi_m.$$
Here
$$\Phi_1=\frac{e}{\epsilon r_1}\Theta(r_1-c_nt-\frac{z_0}{\beta_n}),
\quad\Phi_2=\frac{e}{\epsilon r_2}\Theta(c_nt-r_2-\frac{z_0}{\beta_n}),$$
$$\Phi_m=\Phi_m^{(1)}+\Phi_m^{(2)}+ \Phi_m^{(3)},
\quad  A_z=A_z^{(1)}+A_z^{(2)}+ A_z^{(3)},
\quad\Phi_m^{(i)}=\frac{1}{\epsilon\beta}A_z^{(i)} \eqno(3.3) $$
$$ A_z^{(1)}=\frac{e\beta}{r_m}\Theta(\rho\gamma_n-z-z_0) 
\Theta(\frac{z_0}{\beta_n}+r_2-c_nt) \Theta(c_nt+\frac{z_0}{\beta_n}-r_1),$$
$$ A_z^{(2)}=\frac{e\beta}{r_m} \Theta(z-z_0-\rho\gamma_n) 
\Theta(r_1-c_nt-\frac{z_0}{\beta_n}) \Theta(c_nt-\frac{z_0}{\beta_n}-r_2),$$
$$ A_z^{(3)}=\frac{e\beta}{r_m}\Theta(z_0+\rho\gamma_n-z)
\Theta(z+z_0-\rho\gamma_n)\Theta(c_nt-R_m)\cdot$$
$$[\Theta(r_1-c_nt-\frac{z_0}{\beta_n})+
\Theta(\frac{z_0}{\beta_n}+r_2-c_nt)],$$
(for simplicity we have omitted the $\mu$ factor).

Theta functions
$$\Theta(c_nt+\frac{z_0}{\beta_n}-r_1)\quad {\rm and}
\quad \Theta(r_1-c_nt-\frac{z_0}{\beta_n})$$
define space regions which, correspondingly, have and have not been reached
by the $BS_1$ shock wave. Similarly, theta functions
$$\Theta(c_nt-\frac{z_0}{\beta_n}-r_2) \quad{\rm and} 
\quad\Theta(r_2-c_nt+\frac{z_0}{\beta_n})$$
define space regions which correspondingly have and have not been reached
by the $BS_2$ shock wave. Finally, theta function
$$\Theta(c_nt-R_m)$$
defines space region that has  been reached by the $CSW$.

The potentials $\Phi_1$ and $\Phi_2$ correspond to the electrostatic fields 
of the charge resting at  $z=-z_0$  up to a moment 
$-t_0$ and at $z=z_0$  after
the moment $t_0$ whilst $\Phi_m$ and $A_z$ 
describe the field of a moving charge. 
Schematic representation of the shock waves position at the fixed moment 
of time is shown in Fig. 1. In the space regions 1 and 2 corresponding to 
$z<\rho\gamma_n-z_0$ and $z>\rho\gamma_n+z_0$, resp.,
there are observed only $BS$ shock waves. In the space region 1
(where $A_z^{(1)}\ne 0,\quad A_z^{(2)}=A_z^{(3)}=0$), at the fixed
observation point the $BS_1$ shock wave (defined by $c_nt+z_0/\beta_n=r_1$)
arrives first and $BS_2$ wave (defined by $c_nt-z_0/\beta_n=r_2$)  later.
In the space region 2 (where $A_z^{(2)}\ne 0,\quad A_z^{(1)}=A_z^{(3)}=0$),
these waves arrive in the reverse order. In the space
region 3 (where $A_z^{(3)}\ne 0,\quad A_z^{(1)}=A_z^{(2)}=0$),
defined by $ \rho\gamma_n-z_0<z<\rho\gamma_n+z_0$, there are
$BS_1$, $BS_2$ and $CSW$ shock waves. 
The latter is defined by the equation $c_nt=R_m$.
Before the arrival of the $CSW$ (i.e., for $R_m>c_n t$) 
there is an electrostatic field of a charge which is at rest at $z=-z_0$. 
After the arrival of the last of the BS shock waves there is 
an electrostatic field of a charge which is at rest at $z=z_0$.
The space region, where $\Phi_m$ and $A_z$ (and, therefore, 
the field of a moving charge) differ from zero, lies between 
the $BS_1$ and $BS_2$ shock waves in the regions 1 and 2 and between $CSW$ 
and one of the $BS$ shock waves in the region 3 (for details see Ref. [14]).
Space region 3 in its turn consists  of two subregions 
$3_1$ and $3_2$ defined by the equations 
$ \rho\gamma_n-z_0<z< (\rho^2\gamma_n^2+z_0^2/\beta_n^2)^{1/2}$
and $(\rho^2\gamma_n^2+z_0^2/\beta_n^2)^{1/2}<z< \rho\gamma_n+z_0$, resp.
In the region $3_1$ at first there arrive $CSW$, then $BS_1$ and, finally, 
$BS_2$.
In region $3_2$ two last waves arrive in the reverse order.\\
In brief, $A_z^{(1)}$ and $A_z^{(2)}$ describe the bremsstrahlung in space
regions 1 and 2, resp., while $A_z^{(3)}$ describe bremsstrahlung and
Cherenkov radiation in space region 3.
The polarization vectors of bremsstrahlungs are tangential to the spheres
$BS_1$ and $BS_2$ and lie in the $\phi=const$ plane coinciding with the plane
of Fig.1. They are directed along the unit vectors 
$\vec n_\theta^{(1)}$ and $\vec n_\theta^{(2)}$, resp.
The polarization vector of $CSW$ (directed along $\vec n_m$)
lies on the CSW. It is shown by the solid line in Fig.1 and also lies in the
$\phi=const$ plane. The magnetic field having only the $\phi$ nonvanishing
component is normal to the plane of figure. 
The Poynting vectors defining the direction
of the energy transfer are normal to $BS_1$, $BS_2$ and $CSW$, resp.\\
The Cherenkov radiation in the ($\rho,z$) plane differs
from zero inside the beam of the width $2z_0\sin\theta_c$, 
where $\theta_c$ is the inclination of the beam towards the motion axis 
($\cos\theta_c=1/\beta_n$).
When the charge velocity tends to the velocity of light in medium, the width
of the above beam as well as the inclination angle tend to zero. 
That is, in this case the beam propagates in a nearly forward direction. 
It is essentially
that Cherenkov beam exists for any motion interval $z_0$.

\subsection{Time evolution of the electromagnetic field on the sphere surface}
Consider the distribution of VP (in units $e/R_0$) on the sphere $S_0$ 
of the radius $R_0$ at different moments of time.
There is no EMF on $S_0$ up to a moment $T_n=1-\epsilon_0(1+1/\beta_n)$.
Here $T_n=c_n t/R_0$. In the time interval
$$1-\epsilon_0(1+\frac{1}{\beta_n})\le T_n\le 1-\epsilon_0
(1-\frac{1}{\beta_n})\eqno(3.4)$$
BS radiation begins to fill  the back part of $S_0$ 
corresponding to the angles
$$-1<\cos\theta<\frac{1}{2\epsilon_0}[(T_n+
\frac{\epsilon_0}{\beta_n})^2-1-\epsilon_0^2]\eqno(3.5)$$
(Fig. 2a, curve 1). In the time interval
$$ 1-\epsilon_0(1-\frac{1}{\beta_n})\le T_n
\le [1-(\frac{\epsilon_0}{\beta_n\gamma_n})^2]^{1/2}\eqno(3.6)$$
BS radiation begins to fill the front part of $S_0$ as well:
$$\frac{1}{2\epsilon_0} [1+\epsilon_0^2-(T_n-\frac{\epsilon_0}
{\beta_n })^2]\le\cos\theta\le 1.$$
The illuminated back part of $S_0$ is still given by (3.5) (Fig. 2a, curve 2).
The finite jumps of VP shown in these figures lead to the  $\delta$-type
singularities in Eqs. (3.1) defining BS electromagnetic strengths. 
In the time intervals (3.4) and (3.6) these jumps have a finite height. 
The vector potential is maximal at the angle at which the jump occurs. 
The value of VP  is infinite  at the angles defined by
$$\cos\theta_1=-\frac{\epsilon_0}{\beta_n^2\gamma_n^2}+
\frac{1}{\beta_n}[1-(\frac{\epsilon_0}{\beta_n\gamma_n})^2]^{1/2}
\quad {\rm and}\quad \cos\theta_2=\frac{\epsilon_0}{\beta_n^2\gamma_n^2}+
\frac{1}{\beta_n}[1-(\frac{\epsilon_0}
{\beta_n\gamma_n})^2]^{1/2}.\eqno (3.7) $$
which are reached at the time
$$T_{Ch}=\frac{c_n t_{Ch}}{R_0}=[1-(\frac{\epsilon_0}
{\beta_n\gamma_n})^2]^{1/2}$$
(Fig. 2a, curve 3). At this moment and at these angles the $CSW$ 
intersects $S_0$ first time. 
Or, in other words,  the intersection of $S_0$ by the lines
$z=\rho\gamma_n-z_0$ and $z=\rho\gamma_n+z_0$  (Fig.1) occurs at the angles
$\theta_1$ and $\theta_2$.
At this moment the illuminated front and back parts of $S_0$ are given by
$\theta_1<\theta<\pi$  and $0<\theta<\theta_2$, resp.
Beginning from this moment, the $CSW$ intersects the sphere $S_0$ at the angles defined
by (see Fig. 2b)
$$\cos\theta_{Ch}^{(1)}(T)=\frac{T_n}{\beta_n }-
\frac{1}{\beta_n\gamma_n}(1- T_n^2)^{1/2}\quad {\rm and}\quad
\cos\theta_{Ch}^{(2)}(T) =\frac{T_n}{\beta_n }+
\frac{1}{\beta_n\gamma_n}(1- T_n^2)^{1/2}.$$
The positions of the $BS_1$ and $BS_2$ shock waves are given by
$$\cos\theta_{BS}^{(1)}(T)=\frac{1}{2\epsilon_0}[(T_n+\frac{\epsilon_0}{\beta_n})^2-1-\epsilon_0^2]\quad {\rm
and}\quad\cos\theta_{BS}^{(2)}(T)=\frac{1}{2\epsilon_0} 
[1+\epsilon_0^2-(T_n-\frac{\epsilon_0}{\beta_n})^2],$$
respectively (i.e., the $BS$ shock waves follow after the $CSW$). 
Therefore, at this moment $BS$  fills the angle regions
$$\theta_{BS}^{(1)}(T)\le\theta\le\pi\quad {\rm and}\quad 0
\le\theta\le \theta_{BS}^{(2)}(T)$$
while the VC radiation occupies the angle interval
$$ \theta_{Ch}^{(1)}(T)\le\theta\le\theta_1\quad {\rm and}
\quad \theta_2\le\theta\le\theta_{Ch}^{(2)}(T)$$
Therefore, VC radiation field and $BS$ overlap in the regions
$$\theta_{BS}^{(1)}(T)\le\theta\le\theta_1\quad {\rm and}
\quad \theta_2\le\theta\le \theta_{BS}^{(2)}(T).$$
$BS_1$ and $BS_2$ have finite jumps in this angle interval (Fig. 2b).
The non-illuminated part of $S_0$ is
$$\theta_{Ch}^{(2)}(T)\le\theta\le\theta_{Ch}^{(1)}(T).$$
This lasts up to a moment $T_n =1$ when the Cherenkov shock wave intersects
$S_0$ only once at the point corresponding to the angle  
$\cos\theta=1/\beta_n$ (Fig. 2c).
The positions of the $BS_1$ and $BS_2$ shock waves at this moment ($T_n=1$)
are given by
$$\cos\theta=\frac{1}{\beta_n}-\frac{\epsilon_0}{2 \beta_n^2\gamma_n^2}
\quad {\rm and}\quad
\cos\theta=\frac{1}{\beta_n}+\frac{\epsilon_0}{2 \beta_n^2\gamma_n^2},$$
resp. Again, the jumps of BS waves have finite heights while the Cherenkov
potentials (and field strengths) are infinite at the angle 
$\cos\theta=1/\beta_n$ where $CSW$ intersects $S_0$ \\
After the moment $T_n =1$, $CSW$ leaves  $S_0$.  
However, the Cherenkov post-action still remains (Fig. 3a).
At the subsequent moments of time the $BS_1$ and $BS_2$ shock waves approach
each other. They meet at the moment
$$T_n =[1+(\frac{\epsilon_0}{\beta_n\gamma_n})^2]^{1/2}.\eqno(3.8)$$
at the angle
$$\cos\theta=\frac{1}{\beta_n}[1+(\frac{\epsilon_0}
{\beta_n\gamma_n})^2]^{1/2}.$$
After this moment BS shock waves pass through each other 
and diverge (Fig. 3b).
Now $BS_1$ and $BS_2$ move along the front and back semi-spheres, resp.
There is no EMF on the part of $S_0$  lying between them.
The illuminated parts of $S_0$ are now given by
$$\theta_{BS}^{(2)}(T)\le\theta\le\pi\quad{\rm and}
\quad 0\le\theta\le\theta_{BS}^{(1)}(T)$$
The electromagnetic field is zero inside the angle interval
$$\theta_{BS}^{(1)}(T)\le\theta\le\theta_{BS}^{(2)}(T).$$
After the moment of time (3.8) $BS_1$ and $BS_2$ may occupy the same angular
positions $cos\theta_2$ and $cos\theta_1$ 
like $BS_2 $ and $BS_1$ shown by curve
3 in Fig. 2a. But now their jumps are finite. After the moment
$$T_n=1+\epsilon_0(1-\frac{1}{\beta_n})$$
the front part of $S_0$ begins not to be illuminated (Fig. 3c). At this 
moment the illuminated back part of $S_0$ is given by
$$-1\le\cos\theta\le -1+\frac{2(1+\epsilon_0)}
{\beta_n}-\frac{2\epsilon_0}{\beta_n^2}.$$
In the subsequent time the illuminated part of $S_0$ is given by
$$-1\le\cos\theta\le\frac{1}{2\epsilon_0} 
[1+\epsilon_0^2-(T_n-\frac{\epsilon_0}{\beta_n })^2]$$.
As time goes, the illuminated part of $S_0$ diminishes. 
Finally , after the moment
$$T_n=1+\epsilon_0(1+\frac{1}{\beta_n})$$
the EMF radiation leaves  the surface of  $S_0$ (and its interior).

We  summarize here main differences between Cherenkov radiation and
bremsstrahlung:\\
On the sphere $S_0$, VC radiation runs over the angular region
$$\theta_2\le\theta\le\theta_1,$$
where $\theta_1$ and $\theta_2$ are defined by Eqs. (3.7).
At each particular moment of time $T_n$ in the interval
$$[1-(\frac{\epsilon_0}{\beta_n\gamma_n})^2]^{1/2}\le T_n\le1$$
the VC electromagnetic potentials and field strengths are infinite at the
angles $\theta_{Ch}^{(1)}(T)$ and $\theta_{Ch}^{(2)}(T)$
at which $CSW$ intersects $S_0$.\\
After the moment $T_n=1$ the  Cherenkov singularity leaves  the sphere
$S_0$, but the Cherenkov post-action still remains. This lasts up to the
moment $T_n =[1+(\epsilon_0/\beta_n\gamma_n)^2]^{1/2}$.  \\
On the other hand, $BS$ runs over  the whole sphere $S_0$ in the time interval
$$1-\epsilon_0(1+\frac{1}{\beta_n})\le T_n\le 1+\epsilon_0
(1+\frac{1}{\beta_n}).$$
The vector potential of $BS$  is infinite only at the angles $\theta_1$
and $\theta_2$ at the particular moment of time 
$T_n=\sqrt{1-\epsilon_0^2/\beta_n^2\gamma_n^2}$
when $CSW$ first time intersects $S_0$.
For other times the VP of $BS$ exhibits  finite jumps in the angle interval
$-\pi\le\theta\le\pi$. The $BS$ electromagnetic field strengths 
(as space-time derivatives of electromagnetic potentials) are infinite
at those angles.  Therefore, Cherenkov singularities of the vector potential
run over the region $\theta_2\le\theta\le\theta_1$ of the sphere $S_0$,
while the $BS$ vector potential is infinite only at the angles  
$\theta_1$ and $\theta_2$ where $BS$ shock waves meet $CSW$.

The following particular cases are of special interest.
For small $\epsilon_0=z_0/R_0$ the Cherenkov singular radiation 
occupies the narrow angular region
$$\frac{1}{\beta_n}-\frac{\epsilon_0}{\beta_n^2\gamma_n^2}
\le\cos\theta\le\frac{1}{\beta_n}+\frac{\epsilon_0}{\beta_n^2\gamma_n^2},$$
while BS is infinite at the boundary points of this interval ( at
$\cos\theta=\frac{1}{\beta_n}\pm\frac{\epsilon_0}
{\beta_n^2\gamma_n^2}$) reached
at the moment $T_n=1-\epsilon_0^2/2\beta_n^2\gamma_n^2$.\\
In the opposite case ($\epsilon_0\approx 1$) 
the singular Cherenkov radiation
field is confined to the angular region
$$\frac{2}{\beta_n^2}-1\le\cos\theta\le 1,$$
while BS has singularities at $\cos\theta=\frac{2}{\beta_n^2}-1,\quad
{\rm and}\quad \cos\theta= 1$ reached at the moment $T_n=1/\beta_n$.\\
When the charge velocity is close to the light velocity in medium 
($\beta_n\approx 1$), one gets:
$$\cos\theta_1\approx\frac{1}{\beta_n}-\frac{\epsilon_0}
{\beta_n^2\gamma_n^2}(1+\frac{1}{2}\epsilon_0)\approx 1,\quad
\cos\theta_2\approx\frac{1}{\beta_n}-\frac{\epsilon_0}
{\beta_n^2\gamma_n^2}(1-\frac{1}{2}\epsilon_0)\approx 1,$$
i.e., there is a narrow Cherenkov beam in the nearly forward direction.

\subsection{Comparison with Tamm's vector potential}
Now we evaluate Tamm's VP
$$A_T=\int\limits_{-\infty}^{\infty}d\omega\exp{(i\omega t)}A_\omega$$
Substituting here $A_\omega$ given by (2.2), we get in the absence 
of dispersion
$$A_T=\frac{e}{R_0 n|\cos\theta-1/\beta_n|}
\Theta(|\cos\theta-1/\beta_n|-|T_n-1|/\epsilon_0).\eqno(3.9)$$
This VP may be also obtained  from $A_z$ given by (3.3) 
if we leave in it the terms
$A_z^{(1)}$ and $A_z^{(2)}$ describing BS in the regions 1 and 2 (see Fig.1)
(with omitting $z_0$ in the factors $\Theta(\rho\gamma_n-z-z_0)$
and $ \Theta(z-z_0-\rho\gamma_n)$ entering into them) and 
drop  the term $A_z^{(3)}$ which is responsible 
(as we have learned from the previous section) for the  $BS$ and
$VC$ radiation  in region 3 and which describes a very thin Cherenkov beam 
in the limit $\epsilon_0\to 0$.
It is seen at once that $A_z$ is infinite only at
$$T_n=1,\quad \cos\theta=1/\beta_n.\eqno(3.10)$$
This may be compared with the exact consideration of the previous section
which shows that the $BS$ part of $A_z$ is infinite at the moment
$$T_{Ch}=\frac{c_n t_{Ch}}{R_0}=[1-(\frac{\epsilon_0}
{\beta_n\gamma_n})^2]^{1/2}\eqno(3.11) $$
at the angles $\cos\theta_1$ and $\cos\theta_2$ defined by (3.7).
It is seen, $\cos\theta_1$ and $\cos\theta_2$ defined by (3.7) 
and $T_{Ch}$ given
by (3.11) are transformed into $\cos\theta$ and $T_n$ given by (3.10) 
in the limit $\epsilon_0\to 0$.
Due to the dropping of the $A_z^{(3)}$ term in (3.3) 
(describing bremsstrahlung and Cherenkov radiation in space region 3) 
and the omission of terms containing $\epsilon_0$ in  
$\cos\theta_1$ and $\cos\theta_2$, $BS_1$ and $BS_2$
waves have now the common  maximum of the infinite height at the angle
$\cos\theta=1/\beta_n$ where Tamm's approximation fails.

The analysis of (3.9) shows that Tamm's VP is distributed over $S_0$ in the
following way. There is no EMF of the moving charge up to the moment
$T_n=1-\epsilon_0(1+1/\beta_n)$. For
$$1-\epsilon_0(1+\frac{1}{\beta_n})<T_n<1-\epsilon_0(1-\frac{1}{\beta_n})$$
EMF fills only the back part of $S_0$
$$-1<\cos\theta<\frac{1}{\beta_n}-\frac{1}{\epsilon_0}(1-T_n)$$
(Fig. 4a, curve 1). In the time interval
$$1-\epsilon_0(1-\frac{1}{\beta_n})<T_n< 1+\epsilon_0(1-\frac{1}{\beta_n}) $$
the illuminated parts of $S_0$ are given by
$$-1<\cos\theta<\frac{1}{\beta_n}-\frac{1}{\epsilon_0}(1-T_n)
\quad {\rm and}\quad
\frac{1}{\beta_n}+\frac{1}{\epsilon_0}(1-T_n)<\cos\theta<1$$
(Fig. 4a, curves 2 and 3).
The jumps of the $BS_1$ and $BS_2$ shock waves are finite. As $T_n$
tends to 1, the $BS_1$ and $BS_2$ shock waves approach each other and 
fuse at $T_n=1$.
Tamm' VP is infinite at this moment at the angle 
$\cos\theta=1/\beta_n$ (Fig. 4b).
For
$$1<T_n<1+\epsilon_0(1-\frac{1}{\beta_n})$$
the $BS$ shock waves pass through each other and begin to diverge, 
$BS_1$ and $BS_2$
filling the front and back parts of $S_0$, resp. (Fig. 4c):
$$\frac{1}{\beta_n}+\frac{1}{\epsilon_0}(T_n-1)<\cos\theta<1 
\quad (BS_1) \quad {\rm and}$$
$$-1<\cos\theta<\frac{1}{\beta_n}-\frac{1}{\epsilon_0}(T_n-1)\quad (BS_2).$$
For larger  times
$$1+\epsilon_0(1-\frac{1}{\beta_n})<T_n< 1+\epsilon_0(1+\frac{1}{\beta_n}) $$
only back part of $S_0$ is illuminated:
$$-1<\cos\theta<\frac{1}{\beta_n}-\frac{1}{\epsilon_0}(T_n-1)\quad (BS_2).$$
Finally, for $T_n>1+\epsilon_0(1+1/\beta_n)$ 
there is no radiation field on $S_0$ and inside it.\\
It is seen that the behaviour of exact and approximate Tamm's potentials 
is very alike in the space regions 1 and 2 where Cherenkov radiation 
is absent and differs appreciably in the space region 3 where it exists.
Roughly speaking, Tamm's vector potential (3.9) describing evolution 
of BS shock waves  in the absence of $CSW$ imitates the latter 
in the neighborhood of
$\cos\theta=1/\beta_n$ where, as we know from sect. (3.2), 
Tamm's approximate VP is not correct. \\
This complication is absent if the
charge velocity $\beta$ is less than light velocity in medium $\beta_c$.
In this case one the exact VP is (see [14]):
$$A_z=\frac{e\beta\mu}{r_m}\Theta[c_n(t+t_0)-r_1] \Theta[r_2-c_n(t-t_0)],$$
while Tamm's VP $A_T$ is still given by (3.9). The results of calculations for
$\beta=0.7,\quad  \beta_c\approx 0.75$ are
presented in Fig. 5.  We see on it the exact and Tamm's VPs for three typical
times: $T=1.26;\quad T=1.334$ and $T=1.4$. In general, EMF distribution on the
sphere surface is as follows. There is no field on $S_0$ up to some moment
of time. Later, only back part of $S_0$ is illuminated (see Fig. 5a). 
In the subsequent times the EMF fills the whole sphere (Fig. 5 b). 
After some moment, the EMF again
fills only the back part of $S_0$ (Fig. 5c). Finally, EMF leaves $S_0$.

Now we analyze the behaviour of Tamm's VP for small and large motion
intervals $z_0$. For small $\epsilon_0=z_0/R_0$ it follows from (3.9) that
$A_z=0$ except for the moment $T_n=1$ when
$$A_z=\frac{e}{R_0 n|\frac{1}{\beta_n}-\cos\theta|}.\eqno(3.13)$$
On the other hand, if we pass to the limit $\epsilon_0\to 0$ in Eq.(2.2),
i.e., prior to the integration, then
$$A_\omega\to\frac{e\epsilon_0}{\pi c}\exp(-i\omega R_0/c_n),\quad
A_z\to \frac{e\epsilon_0}{\pi n R_0}\delta(T_n-1),\eqno(3.14)$$
i.e., there is no angular dependence in (3.14). The distinction of (3.14)
from (3.13) is due to the fact that integration takes place for all $\omega$
in the interval ($-\infty,+\infty$). 
For large $\omega$ the condition $\omega z_0/v<<1$
is violated. This means that Eq. (3.13) is more correct.\\
For large $z_0$ one gets from (3.9)
$$A_z=\frac{e}{R_0 n |\frac{1}{\beta_n}-\cos\theta|}.\eqno(3.15)$$
If we take the limit $z_0\to \infty$ in Eq.(2.2), then
$$A_\omega\approx\frac{e\beta}{R_0\omega}\exp(-i\omega R_0/c_n)
\delta(1-\beta_n\cos\theta)\quad
A_z(t)\sim \delta(1-\beta_n\cos\theta).\eqno(3.16)$$
Although Eqs.(3.15) and (3.16) reproduce the position of Cherenkov 
singularity at $\cos\theta=1/\beta_n$, they do not
describe the Cherenkov cone. The reason for this is that Tamm's VP (2.2) is
obtained under the condition $z_0<<R_0$ and, therefore, it is  not legitimate
to take the limit $z_0\to\infty$ in the expressions following from it (and,
in particular, in Eq. (3.9)).\\
On the other hand,  taking the limit $z_0\to\infty$ 
in the exact expression (3.3)
we get the well-known expressions for the electromagnetic potentials 
describing superluminal motion of charge in an infinite medium:
$$A_z=\frac{2e\beta\mu}{r_m}\Theta(vt-z-\rho/\gamma_n),\quad
\Phi=\frac{2e}{\epsilon r_m}\Theta(vt-z-\rho/\gamma_n) .$$
The very fact that Tamm' VP (3.9) is valid both for $\beta<\beta_c$ and
$\beta>\beta_c$ has given rise to the extensive discussion in the physical
literature concerning the discrimination 
between the BS and Cherenkov radiation.
From the facts that: i) Eq.(3.13), following from Tamm's VP (2.2) in the limit
of small $z_0$, does not contain the angular dependence and  ii) this
dependence presents in Eq. (3.15) 
(which differs from zero only for $\beta_n>1$)
following from the same Eq.(2.2) 
in the limit of large $z_0$ it is frequently 
stated (see, e.g., [10,11]) that distinction 
between the Cherenkov radiation and bremsstrahlung disappears 
for $z_0\to 0$ and is maximal for $z_0\to\infty$.

As it follows from our consideration, the physical reason for
this is due to the absence of Cherenkov
radiation in Tamm's VP (3.9). Exact electomagnetic potentials (3.3) and
field strengths (3.1) contain Cherenkov radiation for any $z_0$.
The induced Cherenkov beam being very thin for $z_0\to 0$ 
and broad for large
$z_0$, not in any case can be  reduced to the bremsstrahlung.

This is also confirmed by the consideration of the semi-infinite accelerated
motion of a charged particle in a non-dispersive medium [15]. The arising 
Cherenkov radiation and bremsstrahlung are clearly separated, no ambiguity 
arises in their interpretation.

\section{Space distribution  of Fourier components}
The Fourier transform of the vector potential on the sphere $S_0$ of the radius
$R_0$ is given  by
$$Re A_\omega=\frac{e}{2\pi c}\int\limits_{-\epsilon_0}^{\epsilon_0}
\frac{dz}{Z}\cos[\frac{R_0\omega}{c_n}(\frac{z}{\beta_n}+Z)],$$
$$Im A_\omega=-\frac{e}{2\pi c}\int\limits_{-\epsilon_0}^{\epsilon_0}
\frac{dz}{Z}\sin[\frac{R_0\omega}{c_n}(\frac{z}{\beta_n}+Z)]\eqno(4.1).$$
Here $Z=(1+z^2-2z\cos\theta)^{1/2}$. 
For $z_0<<R_0$ these expressions should be
compared  with the real and imaginary parts of Tamm's approximate VP (2.2):
$$Re A_\omega=\frac{e\beta q}{\pi R_0\omega}
\cos(\frac{\omega R_0}{c_n}),\quad Im A_\omega=-\frac{e\beta q}
{\pi R_0\omega}\sin(\frac{\omega R_0}{c_n}).\eqno(4.2)$$
These quantities are evaluated (in units $e/2\pi c$) for
$$\frac{\omega R_0}{c_n}=100,\quad \beta=0.99,\quad n=1.334,
\quad \epsilon_0=0.1$$
(see Figs. 6 a, b). 
We observe that angular distributions of VPs (4.1) and (4.2)
practically coincide having maxima on the small part of $S_0$ in the
neighborhood of $\cos\theta=1/\beta_n$. It is this minor difference between
(4.1) and (4.2) that is responsible for the Cherenkov radiation 
which is described only by Eq. (4.1). \\
Now we evaluate the angular dependence of VP (4.1) on the sphere $S_0$
for the case when $z_0$ practically coincides with $R_0$ ($\epsilon_0=0.98$).
Other parameters  remain the same. We see ( Fig. 6c) that angular
distribution fills the whole sphere $S_0$. There is no pronounced maximum
in the vicinity of $\cos\theta=1/\beta_n$. \\
We cannot extend these results to larger $z_0$ as the motion interval will
partly lie outside $S_0$. 
To consider a charge motion on an arbitrary finite
interval, we evaluate the distribution of VP on the cylinder 
surface $C$ co-axial with the motion axis. 
Let the radius of this cylinder be $\rho$. Making  the
change of variables  $z=z'+\rho\sinh \chi$ under the sign of integral in
(2.1), one obtains
$$Re A_\omega=\frac{e}{2\pi c}\int\limits_{x_1}^{x_2}
\cos[\frac{\omega\rho}{c}(\frac{z}{\rho\beta}+
\frac{1}{\beta}\sinh\chi+n\cosh\chi)]d\chi,$$
$$Im A_\omega=-\frac{e}{2\pi c}\int\limits_{x_1}^{x_2}
\sin[\frac{\omega\rho}{c}(\frac{z}{\rho\beta}+
\frac{1}{\beta}\sinh\chi+n\cosh\chi)]d\chi,\eqno(4.3)$$
where $x_1=arcsinh\chi_1,\quad x_2=arcsinh\chi_2,
\quad \sinh\chi_1=-(z_0+z)/\rho,\quad \sinh\chi_2=(z_0-z)/\rho$.\\
The distributions of $Re A_\omega$ and $Im A_\omega$ (in units $e/2\pi c$)
on the surface of $C$
as function of $\tilde z=z/\rho$ are shown in Figs 7,8 for different
values of $\epsilon_0=z_0/\rho$ and $\rho$ fixed. 
The calculations were made for
$\beta=0.99$ and $\omega\rho/c=100$. 
We observe that for small $ \epsilon_0$
the electromagnetic field differs from zero only in the vicinity 
$\tilde z= \gamma_n$, which corresponds to $\cos\theta=1/\beta_n$ 
(Fig., 7 a and b). As $\epsilon_0$ increases,
the VP begin to diffuse over the cylinder surface. 
This is illustrated in Figs. 7,c and 8,a where only the real parts of 
$A_\omega$ for $\epsilon_0=1$ and
$\epsilon_0=10$ are presented. Since  the behaviour of $Re A_\omega$ and
$Im A_\omega$ is very much alike 
(Figs. 6, 7 a and b clearly demonstrate this),
we limit ourselves to the consideration of $Re A_\omega$). We observe
the disappearance of pronounced maxima at $\cos\theta=1/\beta_n$. 
For the infinite motion ($z_0\to\infty$) Eqs. (4.3) reduce to
$$Re A_\omega=\frac{e}{2\pi c}\int\limits_{-\infty}^{\infty}
\cos[\frac{\omega\rho}{c}(\frac{z}{\rho\beta}+
\frac{1}{\beta}\sinh\chi+n\cosh\chi)]d\chi,$$
$$Im A_\omega=-\frac{e}{2\pi c}\int\limits_{-\infty}^{\infty}
\sin[\frac{\omega\rho}{c}(\frac{z}{\rho\beta}+
\frac{1}{\beta}\sinh\chi+n\cosh\chi)]d\chi.\eqno(4.4)$$
These expressions can be evaluated in the analytical form (see Appendix)
$$\frac{Re A_\omega}{e/2\pi c}=-\pi [J_0(\frac{\omega\rho}{v\gamma_n})
\sin(\frac{\omega z}{v})+
N_0(\frac{\omega\rho}{v\gamma_n})\cos(\frac{\omega z}{v})] ,$$
$$\frac{Im A_\omega}{e/2\pi c}=  \pi[N_0(\frac{\omega\rho}{v\gamma_n})
\sin(\frac{\omega z}{v})-
J_0(\frac{\omega\rho}{v\gamma_n})\cos(\frac{\omega z}{v})] \eqno(4.5)$$
for $v>c_n$ and
$$\frac{Re A_\omega}{e/2\pi c} =2\cos(\frac{\omega z}{v}) 
K_0(\frac{\rho\omega}{v\gamma_n }),\quad \frac{Im A_\omega}{e/2\pi c}  
=-2\sin(\frac{\omega z}{v}) K_0(\frac{\rho\omega}{v\gamma_n })\eqno(4.6)$$
for $v<c_n$ (remember that $\gamma_n=|1-\beta_n^2|^{-1/2}$). 
We see that for the infinite charge motion the Fourier transform
$A_\omega$ is a pure periodical function of $z$ (and, therefore, of the angle
$\theta$). This assertion does not depend on the $\rho$ and $\omega$ values.
For example, for $\omega\rho/v\gamma_n >>1$ one gets
$$\frac{Re A_\omega}{e/2\pi c}=-\sqrt{\frac{2v\pi\gamma_n}{\rho\omega}}
\sin[\frac{\omega}{v}(z+\frac{\rho}{\gamma_n})-\frac{\pi}{4}],
\quad\frac{Im A_\omega}{e/2\pi c}=-\sqrt{\frac{2v\pi\gamma_n}{\rho\omega}}
\cos[\frac{\omega}{v}(z+\frac{\rho}{\gamma_n})-\frac{\pi}{4}]$$
for $v>c_n$ and
$$\frac{Re A_\omega}{e/2\pi c}=\sqrt{\frac{2v\pi\gamma_n}{\rho\omega}}
\cos(\frac{\omega z}{v})\exp(-\frac{\rho\omega}{v\gamma_n}),
\quad\frac{Im A_\omega}{e/2\pi c}=-\sqrt{\frac{2v\pi\gamma_n}{\rho\omega}}
\sin(\frac{\omega z}{v})\exp(-\frac{\rho\omega}{v\gamma_n})$$
for $v<c_n$.

In Fig. 8b, by comparing the real  part of $A_\omega$
evaluated according to Eq.(4.3) for $\epsilon_0=10$ with the analytical 
expression (4.5) valid for $\epsilon_0\to \infty$ we observe
their perfect agreement on the small interval of cylinder $C$ surface (they
are indistinguishable on the treated interval). 
The same coincidence is valid for $Im A_\omega$.

As it is explicitly stated in [10,11], Tamm's approximate Fourier component  
of VP (2.2) has the $\delta$-type singularity at the Cherenkov angle 
for $z_0\to\infty$  (see Eq.(3.16)) and is independent of angle 
for $z_0\to 0$ (Eq.(3.14)). However, the behaviour of the exact Fourier 
component of VP is  exactly opposite to this behaviour: $A(\omega)$ has 
an isolated maximum for the very small motion intervals
and has infinite number of maxima for $z_0\to\infty$.

The absence of the isolated pronounced maximum  of potentials 
and field strengths at $\cos\theta=1/\beta_n$
for the charge  motion on the finite interval may  qualitatively be 
understood  as follows. We begin with
the exact equations (3.1) and (3.3) for the field strengths and potentials
in the space-time representation. Making inverse Fourier transform from them,
we arrive at Eqs. (4.1)-(4.5) of this section. Now, if the charge motion
takes place on the small space interval, field strengths and potentials
(3.1) and (3.3) have singularities on a rather small  space-time interval
(as the Cherenkov beam is thin in this case).
Therefore, Fourier transforms of (3.1) and (3.3) should be different from zero
in the limited space region. For the charge motion on a large interval
field strengths and potentials (3.1) and (3.3) have singularities in a
larger  space-time domain (as the Cherenkov beam is a rather broad now). 
Consequently, Fourier transforms of (3.1) and (3.3) 
should be different from zero in a larger space region. \\
By comparing (4.4) with (4.5) and (4.6) we recover integrals which,  
to the best of our knowledge, are absent in the mathematical literature 
(see Appendix 1).

\section{Quantum analysis of Tamm's formula}
We turn now to the quantum consideration of Tamm's formula. The usual
approach proceeds as follows [19]. Consider the uniform rectilinear (say,
along the $z$ axis) motion of a point charged particle with the velocity $v$.
The conservation of energy-momentum is written as
$$\vec p=\vec p'+\hbar \vec k,\quad {\cal E}={\cal
E}'+\hbar\omega,\eqno(5.1)$$
where $\vec p$,${\cal E}$  and $\vec p'$,${\cal E}'$ are the 3- momentum
and energy of the initial and final states of the moving charge; $\hbar\vec k$
and $\hbar \omega$ are the 3-momentum and energy
of the emitted photon.  We present (5.1) in the 4-dimensional form
$$p-\hbar k=p',\quad p=(\vec p, {\cal E}/c) .\eqno(5.2)$$
Squaring both sides of this equation and taking into
account that $p^2={p'}^2=-m^2c^2$ ($m$ is the rest mass of a moving charge),
one gets
$$(pk)=\hbar k^2/2,\quad k+(\vec k,\frac{\omega}{c_n}).\eqno(5.3)$$
Or, in a more manifest form
$$\cos\theta_k=\frac{1}{\beta_n}
(1+\frac{n^2-1}{2}\frac{\hbar\omega}{{\cal E}}).\eqno (5.4)$$
Here $\beta_n=v/c_n$, $c_n=c/n$ is the light velocity in medium,
$n$ is its refractive index. 
When deriving (5.4) it was implicitly suggested
that the absolute value of photon 3- momentum and its energy are related
by the Minkowski formula: $|\vec k|=\omega/c_n$.\\
When the energy of the emitted Cherenkov photon is much smaller 
than the energy of a moving charge,
Eq.(5.4)  reduces to
$$\cos \theta_k=1/\beta_n,\eqno(5.5)$$
which can be written in a manifestly covariant form
$$(pk)=0.\eqno(5.6)$$
Up to now we suggested that the emitted  photon has definite energy 
and momentum. According to [20], the  wave function of a photon propagating 
in vacuum is described by the following expression
$$iN\vec e\exp{[i(\vec k\vec r-\omega t)]},\quad (\vec e\vec k)=0,
\quad \vec e^2=1,\eqno(5.7)$$
where $N$ is the real normalization constant and $\vec e$ is the
photon polarization vector lying in the plane passing through $\vec k$ and
$\vec p$:
$$\vec e_\rho=-\cos\theta_k,\quad \vec e_z=\sin\theta_k,
\quad \vec e_\phi=0,\quad (ek)=0.\eqno (5.8)$$
The photon wave function (5.7) identified 
with the classical vector potential is
obtained in the following way. 
We take the positive-frequency part of the
second-quantized vector potential operator and apply it to the coherent state
with the fixed $\vec k$. The eigenvalue of this VP operator is just (5.7).
In the Appendix 2 we show that the gauge invariance permits one 
to present a wave function in
the form having the form of a classical vector potential
$$iN'p_\mu\exp{(ikx)},\quad (pk)=0.\eqno(5.9)$$
where $N'$ is another real constant.
 Now we take into account that photons described by the wave function
(5.7) are created by the axially symmetric current of a moving charge.
According to Glauber ([21], Lecture 3), to obtain VP in the coordinate
representation, one should make superposition of the wave
functions (5.7) by taking into account the relation (5.6) which tells us
that photon is emitted at the Cherenkov angle $\theta_k$ defined by (5.5).
This superposition is given by
$$A_\mu(x)=iN'\int p_\mu\exp{(ikx)}\delta(pk)d^3k/\omega.$$
The factor $1/\omega$ is introduced using the analogy with the
photon wave function in vacuum where it is needed for the relativistic
covariance of $A_\mu$. The expression $p_\mu\delta(pu)$ is (up to
a factor) the Fourier transform of the classical current of the
uniformly moving charge.
This current creates photons in  coherent states which are observed
experimentally. In particular, they are manifested as a classical
electromagnetic radiation. We rewrite $A_\mu$ in a slightly extended form
$$A_\mu=iN'\int p_\mu\exp{[i(\vec k\vec r-\omega t)]}
\delta[\frac{{\cal E}\omega}{c^2}
(1-\beta_n\cos\theta)]\frac{n^3}{c^3}d\phi d\cos\theta \omega
d\omega.\eqno(5.10)$$
Introducing the cylindrical coordinates ($\vec r=\rho\vec n_\rho+z\vec n_z$),
we present $\vec k\vec r$ in the form $$\vec k\vec
r=\frac{\omega}{c_n}[\rho\sin\theta\cos(\phi-\phi_r)+z\cos\theta].$$
Inserting this into (5.10) we get
$$A_\mu(\vec r, t)=iN''\int p_\mu\exp{[i\omega(\frac{z}{c_n}\cos\theta_k
-t)]}\exp{[\frac{i\omega}{c_n}\rho\sin\theta_k
\cos(\phi-\phi_r)]}d\phi d\omega,$$
where $N''$ is the real modified normalization constant and 
$\phi_r$ is the azimuthal angle in the usual space.
Integrating over $\phi$ one gets
$$A_0(\vec r, t)= A_z(\vec r,t)/\beta,\quad
A_z(\vec r, t)=\int\limits_{0}^{\infty}\exp{(-i\omega t)}
A_z(\vec r,\omega) d\omega,$$
where $$A_z(\vec r,\omega)=\frac{2\pi i
N''}{\sin\theta_k}\exp{(\frac{i\omega}{c_n}\cos\theta_k z)}
J_0(\frac{\omega}{c_n}\rho\sin\theta_k).\eqno(5.11)$$
We see that $A_z(\vec r,\omega)$ is the oscillating function of the frequency
$\omega$ without a pronounced $\delta$- type maximum. In the $\vec r,t$
representation $A_z(\vec r,t)$ (and, therefore, photon's wave function)
is singular on the Cherenkov cone $vt-z=\rho/\gamma_n$
$$Re A_z =2\pi N''p_z\int\sin\omega(t-z/v)J_0(\frac{\omega\rho}{c_n}
\sin\theta_k) d\omega=$$
$$=2\pi N''p_z \frac{v}{[(z-vt)^2-\rho^2/\gamma_n^2]^{1/2}}
\Theta((z-vt)^2-\rho^2/\gamma_n^2),$$
$$Im A_z =2\pi N''p_z\int\cos\omega(t-z/v)J_0(\frac{\omega\rho}{c_n}
\sin\theta_k) d\omega=$$
$$=2\pi N''p_z \frac{v}{[\rho^2/\gamma_n^2-(z-vt)^2]^{1/2}} 
\Theta(\rho^2/\gamma_n^2-(z-vt)^2)$$
Despite the fact that the wave function (5.10) satisfies free wave equation 
and does not contain singular Neumann functions $N_0$ 
(needed to satisfy Maxwell equations with a
moving charge current in their r.h.s. ), its real part (which, roughly
speaking, corresponds to the classic electromagnetic potential) properly
describes the main features of the VC radiation.

{\section{Discussion}
So far, our conclusion on the absence of a Cherenkov radiation in Eqs.(2.2)
and (2.3) was proved only for the dispersion-free case (as only in this case 
we have exact solution). 
At this moment we are unable to prove the same result in the general
case with dispersion. We see that Tamm's formulae  describe evolution
and  interference of two BS shock waves emitted at the beginning and at the
end of the charge motion and do not contain the Cherenkov radiation.

Now  the paradoxical results of Refs. [12,13], where the Tamm's
formulae were investigated numerically become understandable. 
Their authors attributed the term $J_{Ch}$
in Eqs. (2.4) to the interference of the bremsstrahlung shock waves emitted
at the moments of instant acceleration and deceleration. Without knowing that
Cherenkov radiation is absent in Tamm's equations (2.2) they concluded that
the Cherenkov radiation is a result of the interference of the above BS shock
waves.We quote them:\\
\begin{quote}
{\footnotesize
"Summing up, one can say that radiation of a charge moving with the light
velocity along the limited section of its path (the Tamm problem) is the
result of interference of two bremsstrahlungs produced in the beginning and
at the end of motion. This is especially clear when the charge moves in
vacuum where the laws of electrodynamics prohibit radiation of a charge
moving with a constant velocity.
In the Tamm problem the constant-velocity charge motion over the distance $l$
between the charge acceleration and stopping moments in the beginning and at
the end of the path  only affects the result of interference but does not 
cause the radiation.

As was shown by Tamm [1] and it follows from our paper the radiation emitted
by the charge moving at a constant velocity over the finite section of the
trajectory $l$ has the same characteristics in the limit $l\to\infty$ 
as the VCR in the Tamm-Frank theory [6]. 
Since the Tamm-Frank theory is a limiting case
of the Tamm theory, one can consider the same conclusion 
is valid for it as well.

Noteworthy is that already in 1939 Vavilov [10] expressed his opinion that
deceleration of the electrons  is the most probable reason for the glow
observed in Cerenkov's experiments".

}
\end{quote}

(We left the numeration of references in this citation the same as it was in
Ref. [12]). We agree with the authors of [12,13] that Tamm's approximate 
formulae (2.2) and (2.3) can be interpreted as the interference between  
two BS waves. This is due to the fact that Tamm's  formulae do not describe 
the Cherenkov radiation properly. 
On the other hand, exact formulae found in [14]  contain
both the Cherenkov radiation and bremsstrahlung and cannot be reduced to the
interference of two BS waves.          \\
Further, we insist that Eq.(1.2) defining the field strength maxima 
in the Fourier representation is valid when the point charge moves 
with the velocity $v>c_n$  on the finite space interval
small compared with the  radius $R_0$ of the observation sphere ($z_0<<R_0$). 
When the value of $z_0$ is compared or larger than $R_0$, 
the pronounced maximum of the Fourier
transforms of the field strengths at the angle $\cos\theta=1/\beta_n$
disappears. Instead, many maxima of the same amplitude  distributed over
the finite region of space arise. In particular, for the infinite charge
motion the above mentioned  Fourier transforms are highly oscillating functions
of space variables distributed over the whole space. This contrasts with the
qualitative analysis of Tamm's approximate  problem given in [10,11] where
the absence of pronounced Cherenkov's radiation maximum and its presence
have been predicted for small and large motion intervals, resp. As it was
shown in sections 3 and 4, this is due to approximations under which Tamm's
electromagnetic potentials and field strengths were obtained.

It follows from the present consideration that Eq. (1.2) 
(relating to the particular Fourier component) cannot be used for 
the identification of the Cherenkov radiation
for large motion intervals. \\
However,  in the usual space-time  representation field strengths
in the absence of dispersion have a singularity  at the angle 
$\cos\theta=1/\beta_n$.
When the dispersion is taken into account, many maxima in the angular
distribution of field strengths 
(in the usual space-time representation) appear,
but the main maximum is at the same position 
where the Cherenkov singularity lies
in  the absence of dispersion ([8]).

It should be noted that doubts on the validity of Tamm's formula (1.2) 
for the maximum of Fourier components were earlier pointed out  
by D.V Skobeltzyne [22] on the grounds entirely different from ours. 
We mean the so-called Abragam-Minkowski
controversy between the photon energy and its momentum.

\section*{Appendix 1}
We start from the  Green function expansion in the cylindrical coordinates
$$G_\omega(\vec r,\vec r')=-\frac{1}{4\pi}
\frac{\exp(-ik_n|\vec r-\vec r'|}{|\vec r-\vec r'| }=$$
$$-\sum\limits_{m=0}^{\infty}\epsilon_m\cos m(\phi-\phi')
\{\frac{1}{4\pi i}\int\limits_{-k_n}^{k_n}dk_z
\exp[k_z(z-z')]{G_m}^{(1)}(\rho,\rho')+$$
$$+\frac{1}{2\pi^2}(\int\limits_{-\infty}^{-k_n}+
\int\limits_{k_n}^{\infty})dk_z\exp[k_z(z-z')]{G_m}^{(2)}(\rho,\rho')\},$$
where $\epsilon_m =1/(1+\delta_{m0})$,
$${G_m}^{(1)}(\rho_<,\rho_>)=J_m(\sqrt{k_n^2-k_z^2}\rho_<)
H_m^{(2)}(\sqrt{k_n^2-k_z^2}\rho_>),$$
$${G_m}^{(2)}(\rho_<,\rho_>)=I_m(\sqrt{k_z^2-k_n^2}\rho_<)
K_m(\sqrt{k_z^2-k_n^2}\rho_>).$$
The Fourier component of VP satisfies the equation
$$(\Delta + k_n^2)A_\omega=-\frac{4\pi}{c}j_\omega,\eqno (A1.1)$$
where $k_n=\omega/c_n>0$ and $j_\omega=\delta(x)\delta(y)
\exp(-i\omega z/v)/2\pi$
The solution of (A1.1) is given by
$$A_\omega=\frac{1}{c} \int G_\omega(\vec r,\vec r')j_\omega(\vec r') dV'=$$
$$=-i\pi\exp(-i\omega z/v) H_0^{(2)}(\frac{\omega\rho}{v}\sqrt{\beta_n^2-1})$$
for $\beta_n>1$ and
$$=2\exp(-i\omega z/v)K_0(\frac{\omega\rho}{v}\sqrt{1-\beta_n^2})$$
for $\beta_n<1$. Separating the real and imaginary parts, we arrive at (4.5).
Equating (4.4) and (4.5) and collecting terms at $\sin(\omega z/v)$ and
$\cos(\omega z/v)$, we get the integrals
$$\int\limits_{0}^{\infty}\cos(\frac{\omega\rho}{v}
\sinh\chi)\sin(\frac{\omega\rho}{c_n}\cosh\chi)d\chi=$$
$$=\int\limits_{0}^{\infty}\cos(\frac{\omega\rho}{v}x)
\sin(\frac{\omega\rho}{c_n}\sqrt{x^2+1})\frac{dx}{\sqrt{x^2+1}}=
\int\limits_{1}^{\infty}\cos(\frac{\omega\rho}{v}\sqrt{x^2-1})
\sin(\frac{\omega\rho}{c_n}x)\frac{dx}{\sqrt{x^2-1}}=$$
$$=\frac{\pi}{2}J_0(\frac{\omega\rho}{v}\sqrt{\beta_n^2-1})\eqno(A1.2)$$
for $v>c_n$ and $=0$ for $v<c_n$.
$$\int\limits_{0}^{\infty}\cos(\frac{\omega\rho}{v}\sinh\chi)
\cos(\frac{\omega\rho}{c_n}\cosh\chi)d\chi=$$
$$=\int\limits_{0}^{\infty}\cos(\frac{\omega\rho}{v}x)
\cos(\frac{\omega\rho}{c_n}\sqrt{x^2+1})\frac{dx}{\sqrt{x^2+1}}=
\int\limits_{1}^{\infty}\cos(\frac{\omega\rho}{v}\sqrt{x^2-1})
\cos(\frac{\omega\rho}{c_n}x)\frac{dx}{\sqrt{x^2-1}}=$$
$$=-\frac{\pi}{2}N_0(\frac{\omega\rho}{v}\sqrt{\beta_n^2-1})\eqno(A1.3)$$
for $v>c_n$ and $=K_0(\frac{\omega\rho}{v}\sqrt{1-\beta_n^2})$ for $v<c_n$.
Here $\beta_n=v/c_n$.\\
As we have mentioned, we did not find these integrals in the available
mathematical literature. In the limit cases these integrals pass into the
tabular ones. 
For example, in the limit $v\to\infty$ Eqs. (A1.2) and  (A1.3) are
transformed into
$$\int\limits_{0}^{\infty}\sin(\frac{\omega\rho}{c_n}\cosh\chi)d\chi=
\frac{\pi}{2}J_0(\frac{\omega\rho}{c_n})\quad {\rm and} \quad
\int\limits_{0}^{\infty}\cos(\frac{\omega\rho}{c_n}\cosh\chi)d\chi=
-\frac{\pi}{2}N_0(\frac{\omega\rho}{c_n}),$$
while Eq. (A1.3) in the limit $c_n\to \infty$ goes into
$$\int\limits_{0}^{\infty}\cos(\frac{\omega\rho}{v}
\sinh\chi)d\chi=K_0(\frac{\omega\rho}{v}).$$

\section*{Appendix 2}

\subsection*{Choice of polarization vector}
The electromagnetic potentials satisfy the following equations
$$(\Delta-\frac{1}{c_n^2}\frac{\partial^2}{\partial
t^2})\vec A=-\frac{4\pi\mu}{c}\vec j,\quad
(\Delta-\frac{1}{c_n^2}\frac{\partial^2}{\partial t^2})\Phi
=-\frac{4\pi}{\epsilon}\rho,$$
$$div\vec A+\frac{\varepsilon\mu}{c}\frac{\partial\Phi}{\partial t}=0.$$
We apply the gauge transformation
$$\vec A\to\vec  A'=\vec A+\nabla \chi,\quad \Phi\to\Phi'=
\Phi-\frac{1}{c}\dot\chi.$$
to the vector potential (5.7) which plays the role of the photon wave
function.  We choose the generating function $\chi$  in the form
$$\chi=\alpha\exp{[i(\vec k\vec r-\omega t)]},$$
where $\alpha$ will be determined later. Thus,
$$\vec A'=(N\vec e+i\alpha\vec k)\exp{[i(\vec k\vec r-\omega t)]} ,
\quad \Phi'=\frac{i\omega\alpha}{c}\exp{[i(\vec k\vec r-\omega t)]} ,$$
where $\vec e$ is given by (5.8). We require the disappearance of
the $\rho$ component of $\vec A'$. This fixes $\alpha$:
$$\alpha=\frac{N}{ik}\cot\theta_k.$$
The nonvanishing components of $\vec A'$ are given by
$$A_z'=\frac{N}{\sin\theta_k}\exp{[i(\vec k\vec r-\omega t)]} ,\quad
A_0'=\frac{N}{n}\cot\theta_k \exp{[i(\vec k\vec r-\omega t)]}.$$
It is easy to see that $A_z'=\beta A_0'$. This completes the proof of (5.9).

\newpage
\begin{figure}[h]
%\vspace*{-15mm}
\begin{center}
\includegraphics[angle=-90,width=90mm]{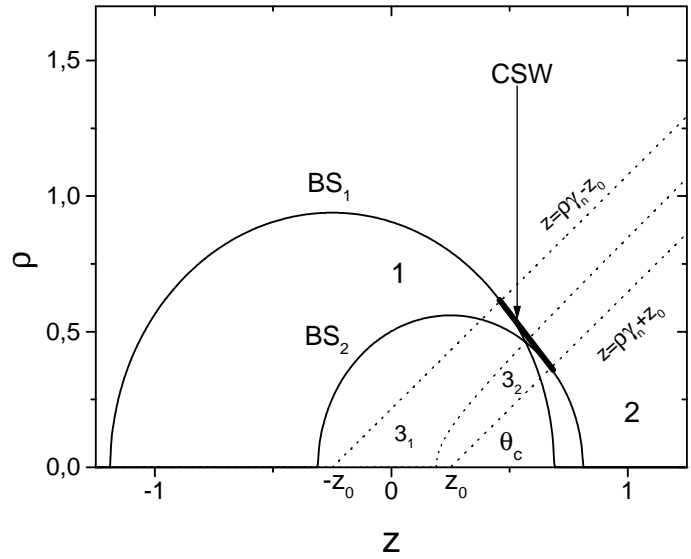}     %fig1
\end{center}
\vspace*{-3mm}
\caption{Position of shock waves at the fixed moment of time
for $\beta=0.99$ and $\beta_c=0.75$.
BS$_1$ and BS$_2$ are bremsstrahlung shock waves emitted
at the points $\mp z_0$ of the $z$ axis.
The solid segment between the lines $z=\rho\gamma_n-z_0$ and
$z=\rho\gamma_n+z_0$ is the \~Cerenkov shock wave (CSW).
The inclination angle of the \~Cerenkov beam and its width are
$\cos\theta_c=1/\beta_c$ and $2z_0\sqrt{1-\beta_n^{-2}}$ resp.}
\end{figure}
\hspace*{-30mm}
\begin{figure}[h]                       %fig2
\includegraphics[height=75mm]{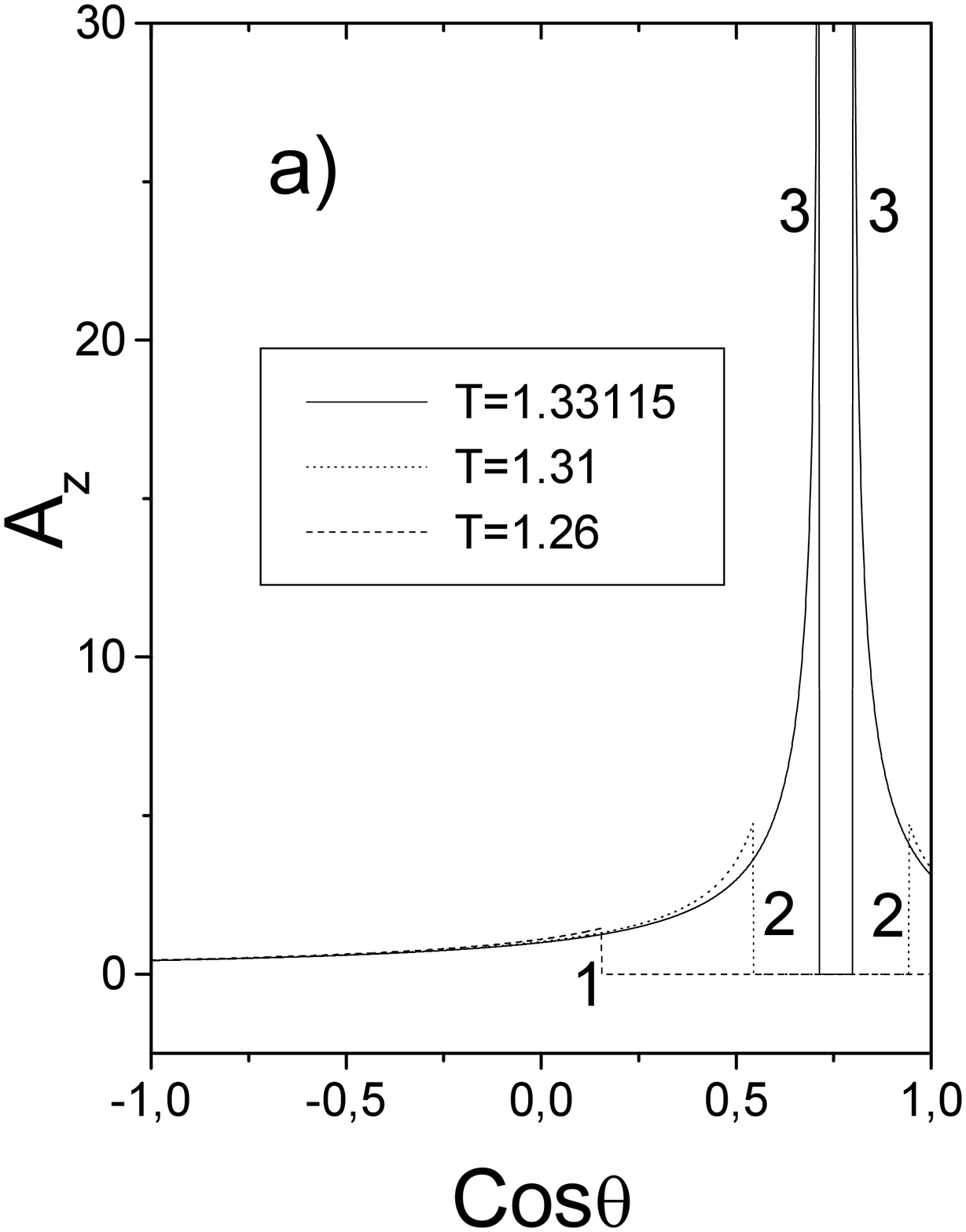}
\includegraphics[height=74mm]{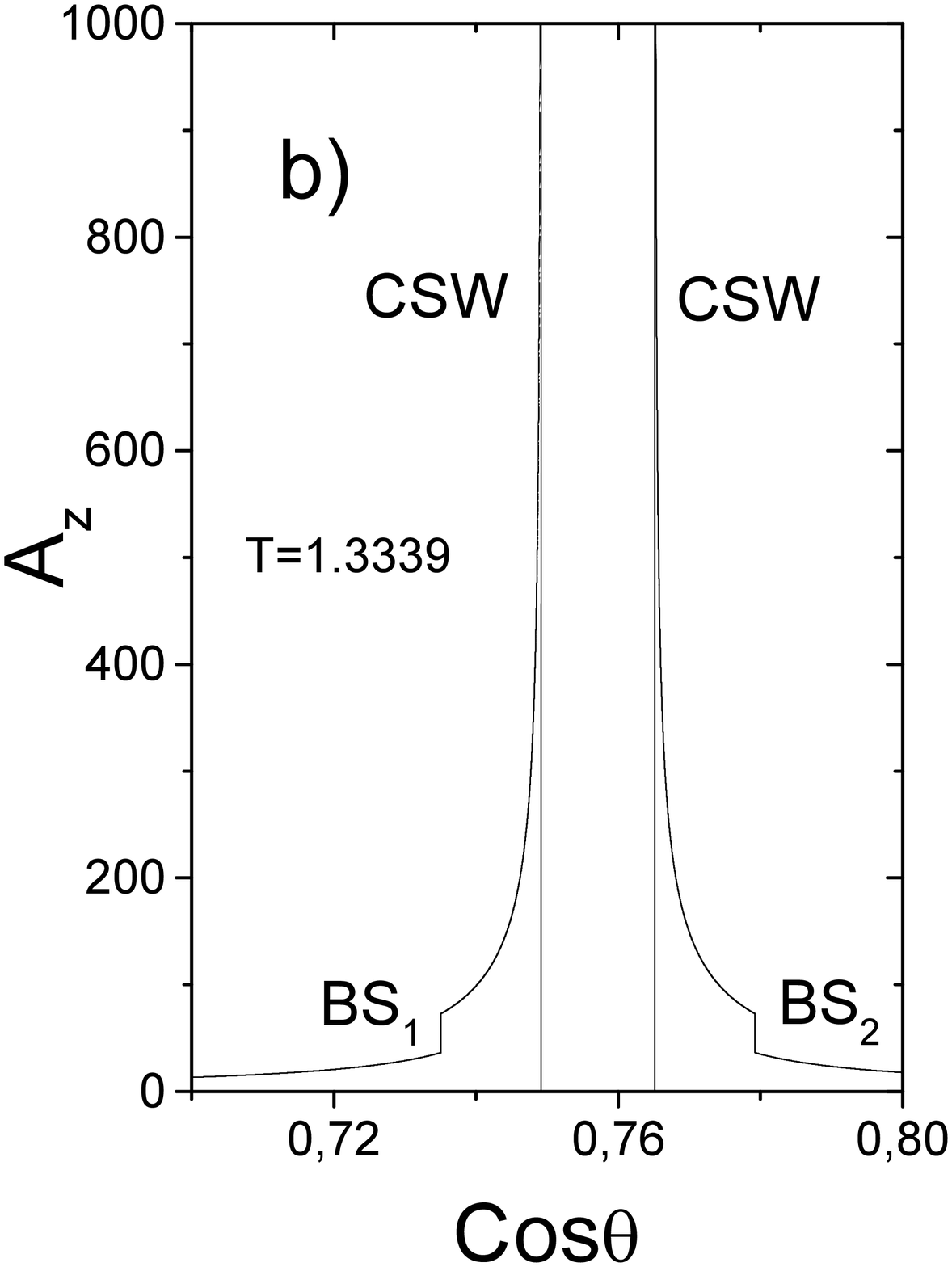}
\includegraphics[height=74mm]{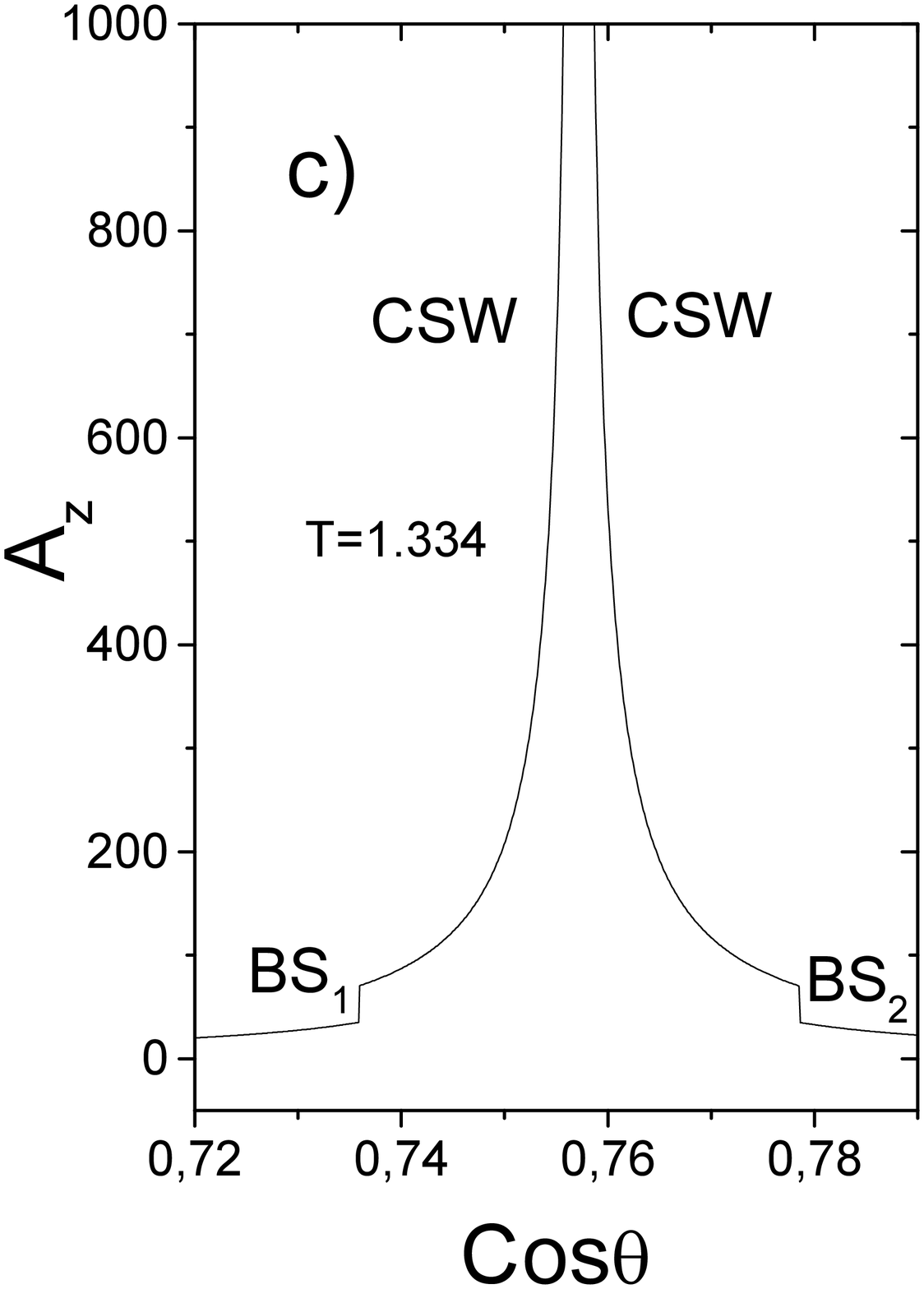}
%\begin{center}
\caption{Time evolution of VP shock waves on the surface of the sphere
S$_0$. $A_z$ is in units $e/R_0$, time $T=ct/R_0$.\\
a) For small times the BS shock wave occupies only back part of
S$_0$ (curve 1). For larger times the BS shock wave begin to fill
the front part of S$_0$ as well (curve 2).
The jumps of BS shock waves are finite. The jump becomes
infinite when the BS shock wave meets CSW (curve 3).\\
b) The amplitude of \~Cerenkov's shock wave is infinite
while BS shock waves exhibit finite jumps.\\
c) Position of CSW and BS shock waves at the moment
when CSW touches the sphere S$_0$ only at one point.}
%\end{center}
\vspace*{-40mm}
\end{figure}

\newpage
\begin{figure}[h]                       %fig.3
%\vspace*{-20mm}
\includegraphics[height=74mm]{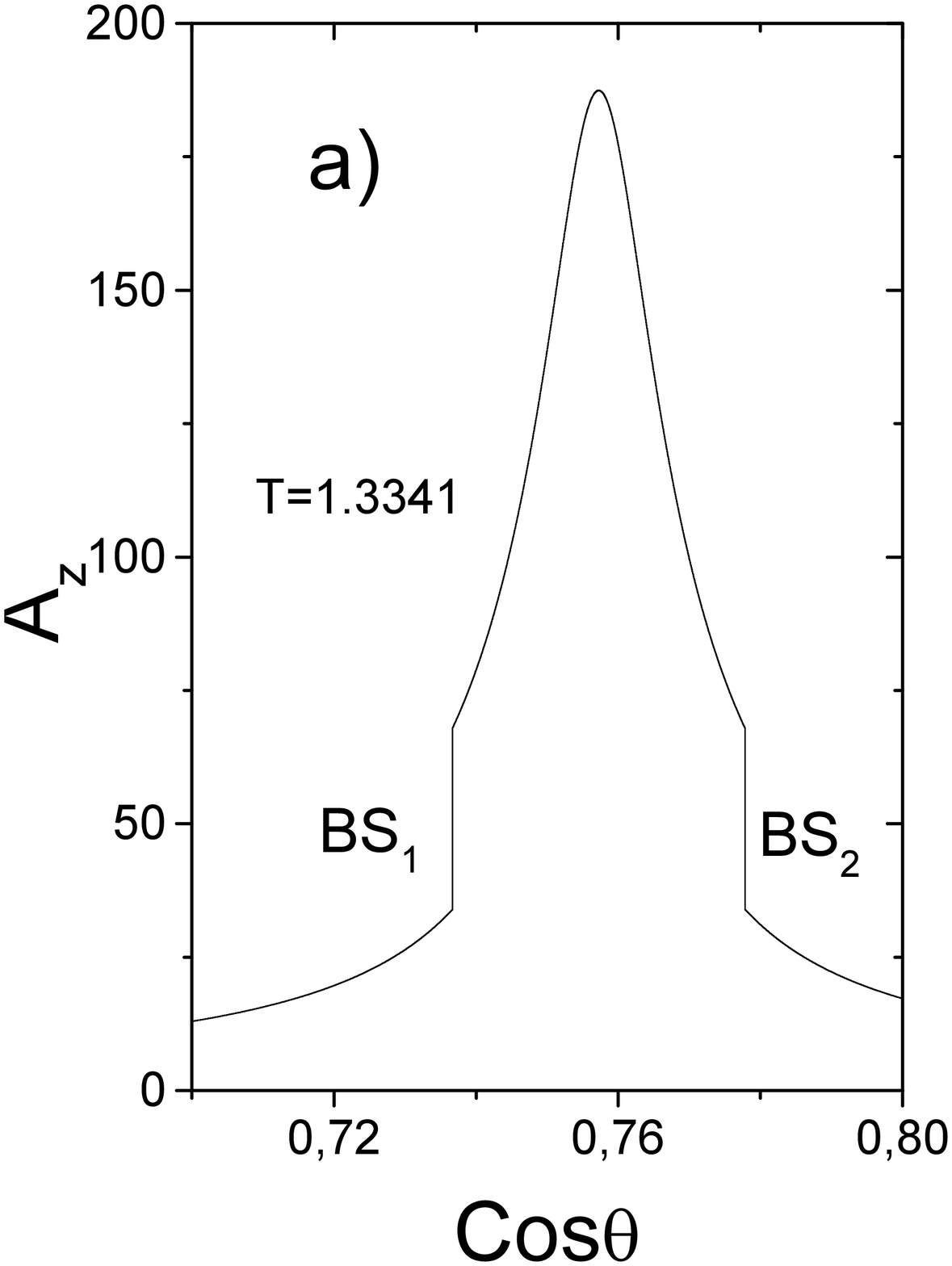}
\includegraphics[height=74mm]{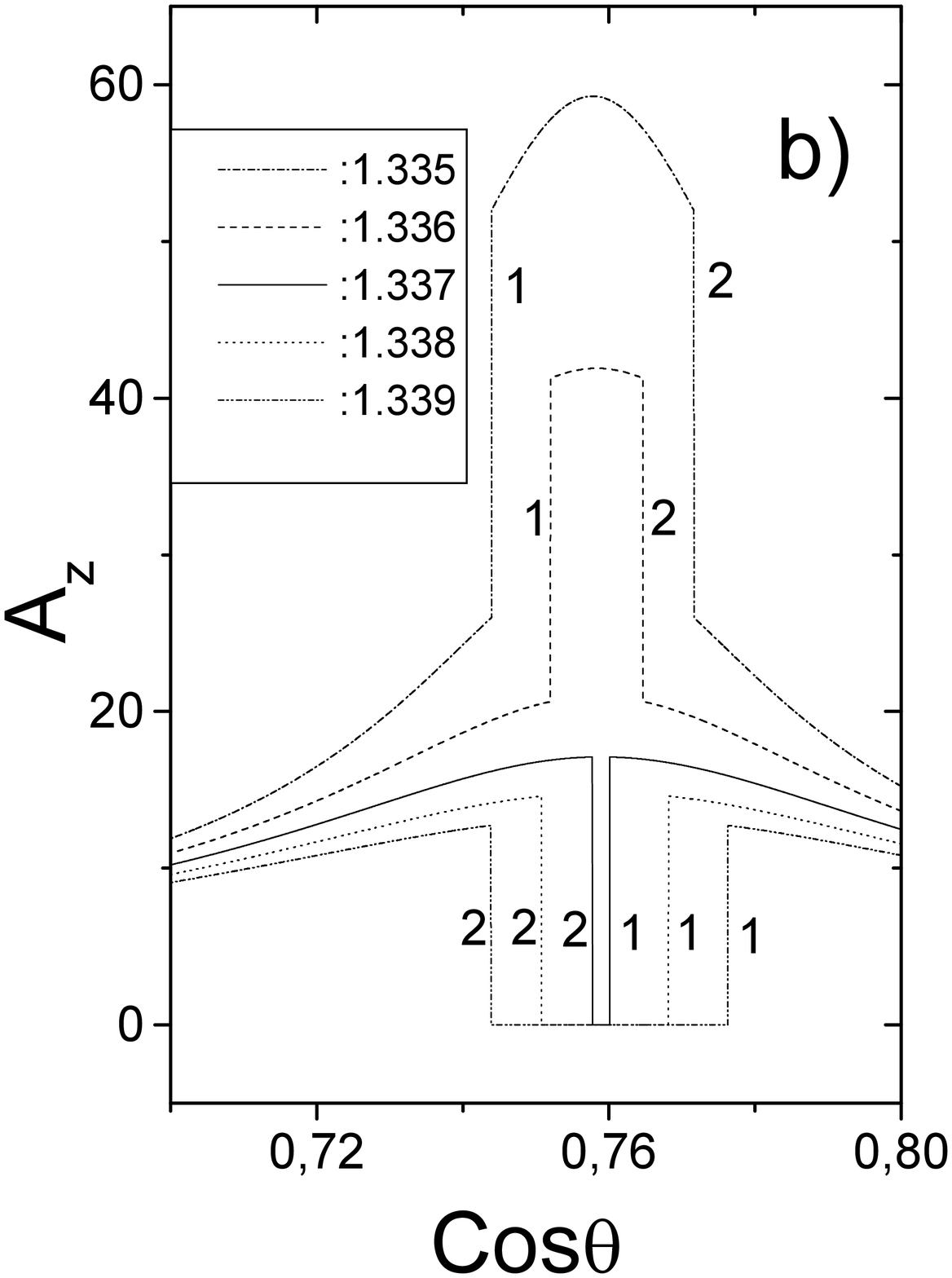}
\includegraphics[height=74mm]{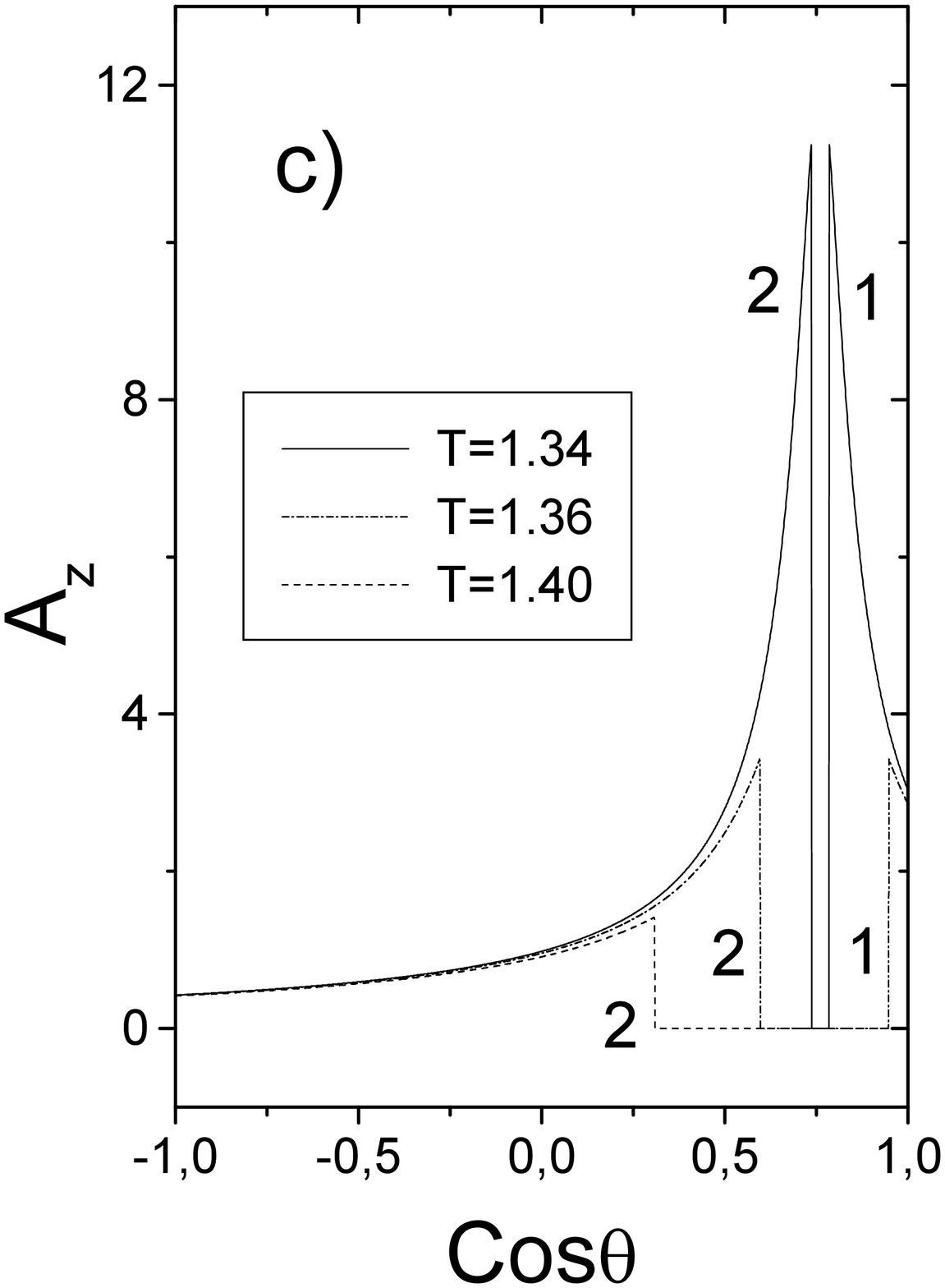}
\caption{Further time evolution of BS shock waves on the surface
of the sphere S$_0$.\\
a) The \~Cerenkov post-action and BS shock waves after
the moment when CSW has left S$_0$.\\
b)  BS shock waves approach and pass through each other
leaving after themselves the zero electromagnetic field.
Numbers 1 and 2 mean BS$_1$ and BS$_2$ shock waves, resp.\\
c) After some moment BS shock wave begin to fill
only the back part of S$_0$.
Numbers 1 and 2 mean BS$_1$ and BS$_2$ shock waves, resp.}
\end{figure}

\begin{figure}[h]               %fig.4
%\vspace*{-8mm}
%\hspace*{-20mm}
\includegraphics[height=74mm]{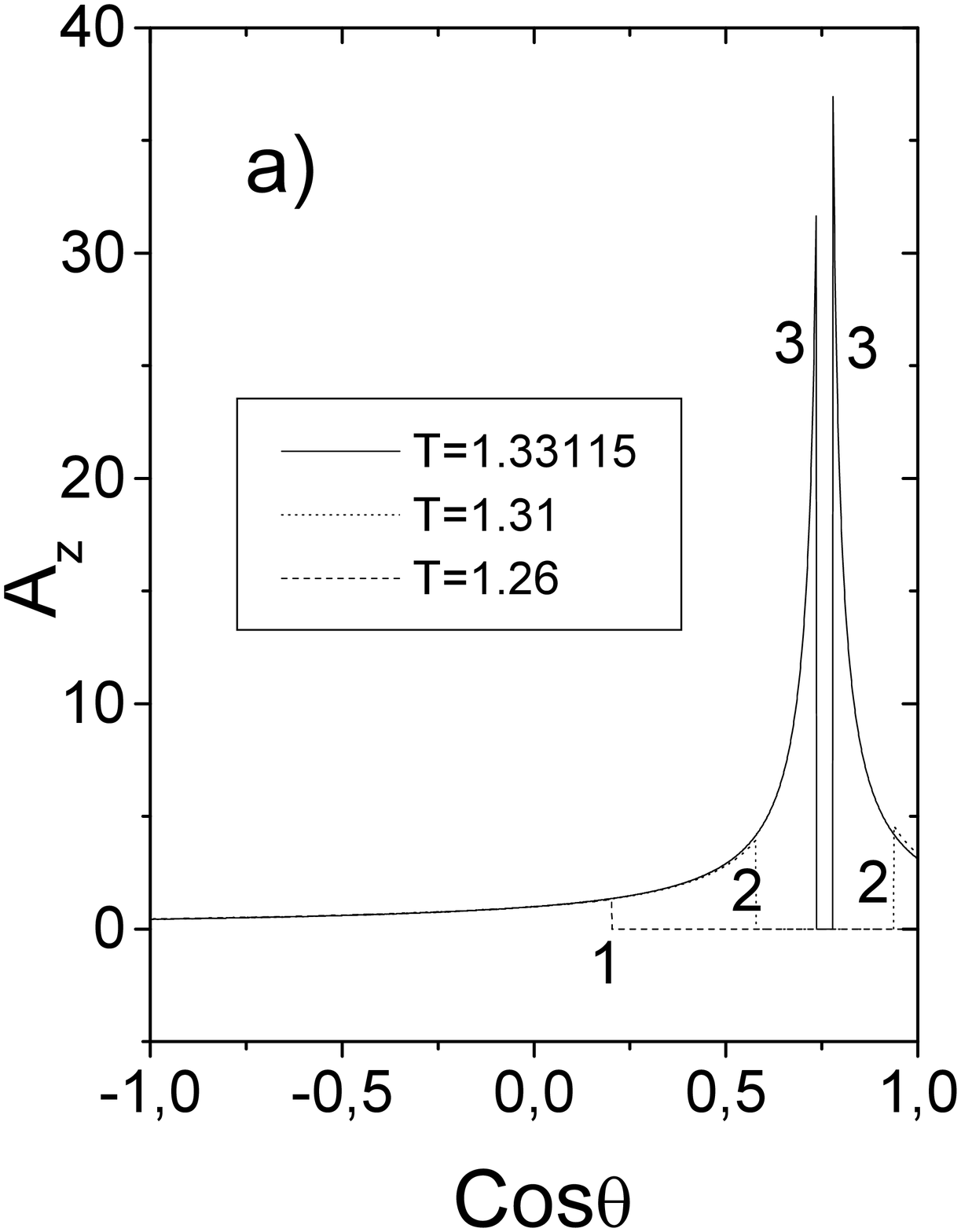}
\includegraphics[height=74mm]{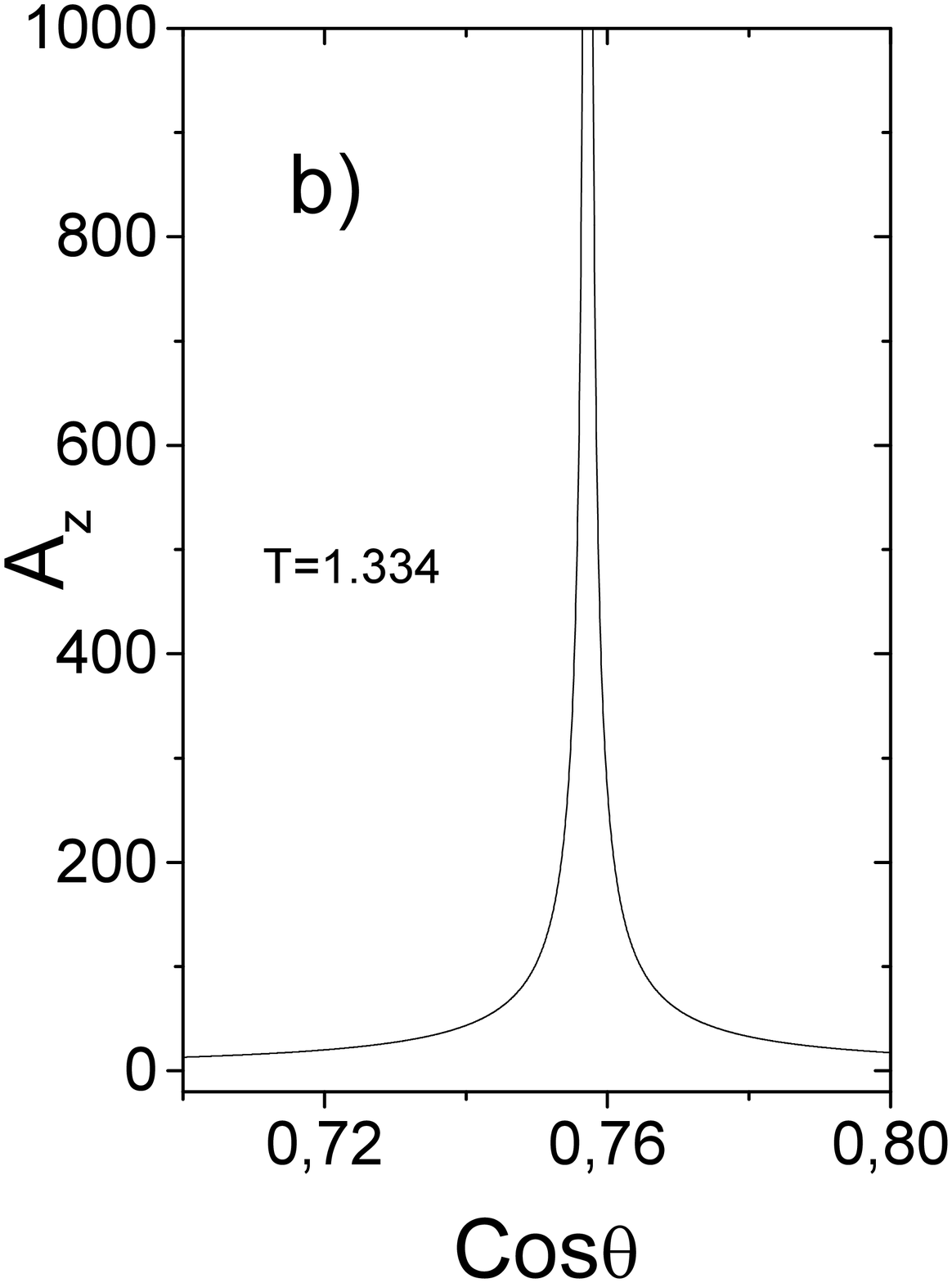}
\includegraphics[height=74mm]{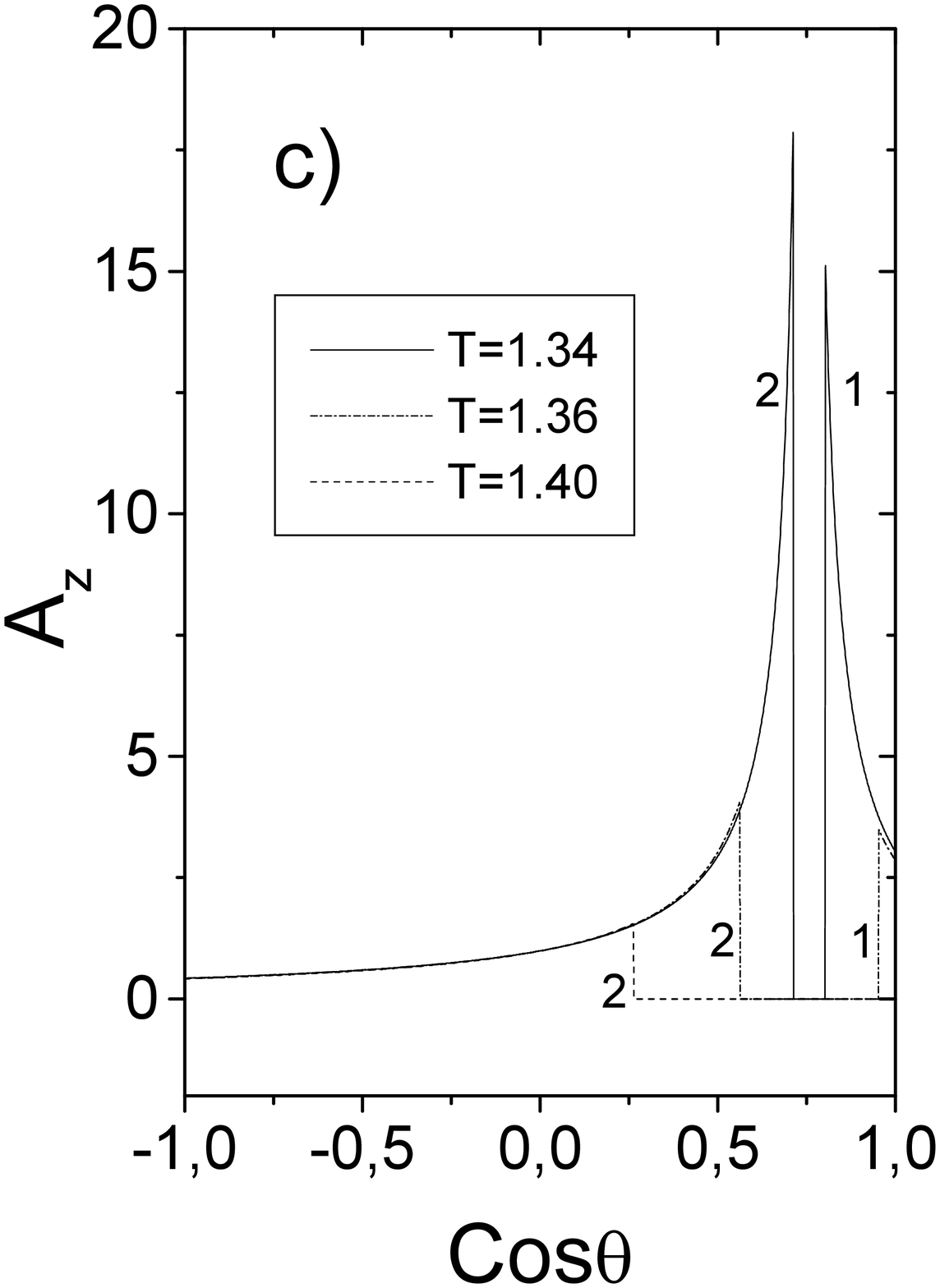}
%\vspace*{-2mm}
\caption{Time evolution of BS shock waves according to
Tamm's approximate picture.\\
a) The jumps of BS shock waves are finite.
After some moment BS shock
waves fill both the back and front parts of S$_0$ (curves 2 and
3).\\
b) Position of the BS shock wave
at the moment when its jump is infinite.\\
c) BS shock waves pass through each other and diverge leaving
after themselves the zero EMF.
After some moment BS shock waves fill
only the back part of S$_0$.
Numbers $1$ and $2$ mean BS$_1$ and BS$_2$ shock waves, resp.}
\vspace*{-45mm}
\end{figure}

\newpage
\begin{figure}[h]                       %fig.5
%\vspace*{-20mm}
\includegraphics[height=74mm]{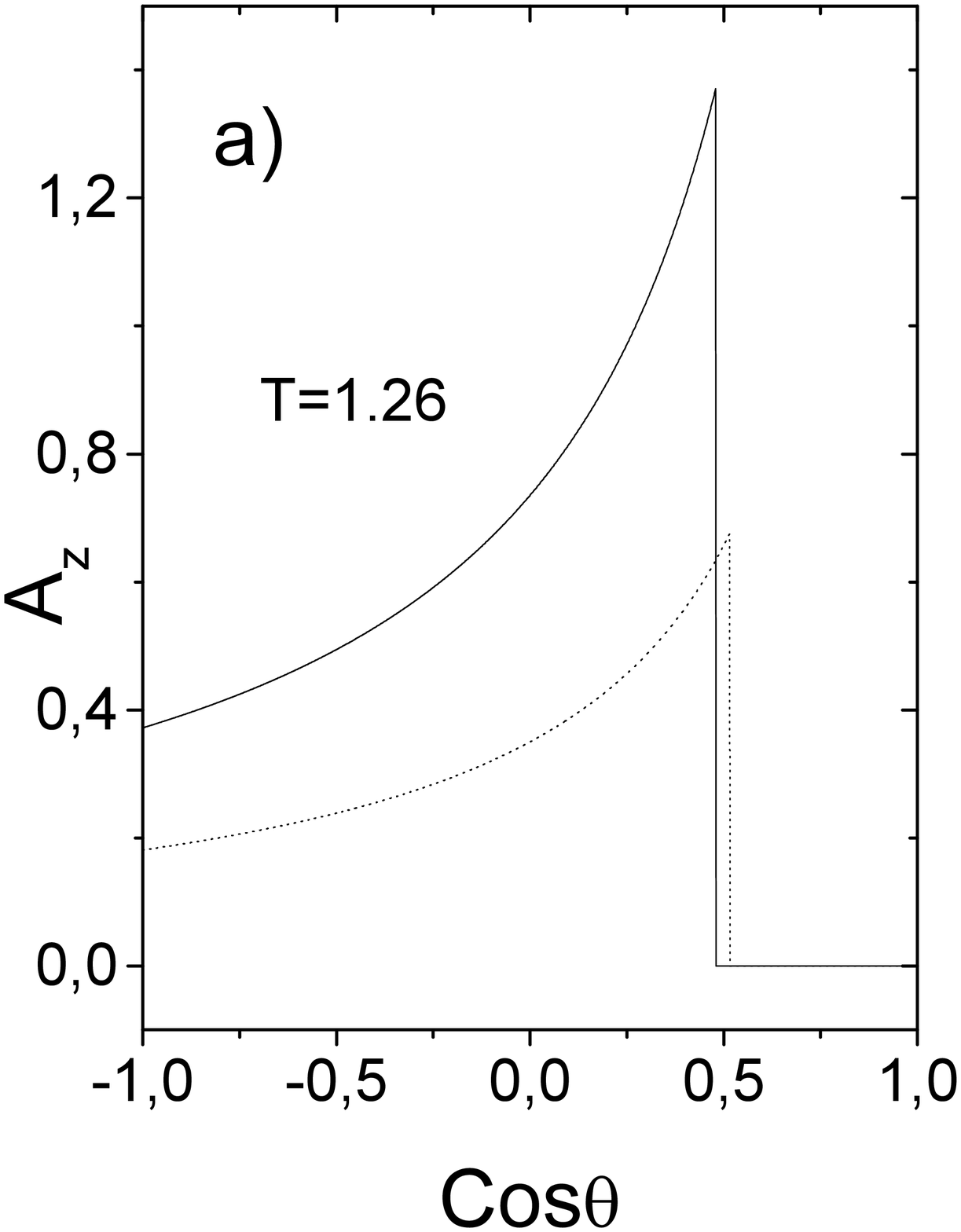}
\includegraphics[height=74mm]{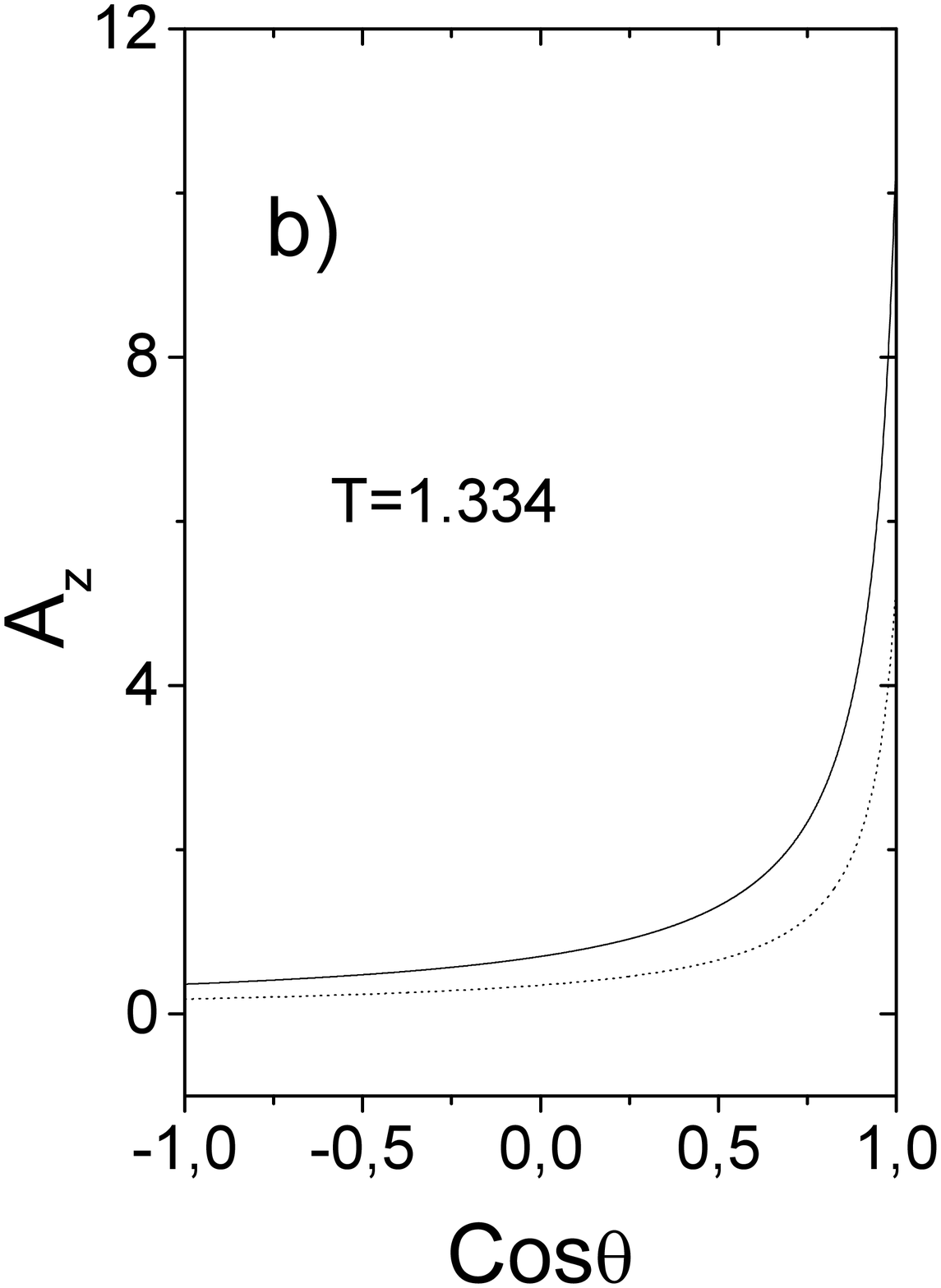}
\includegraphics[height=74mm]{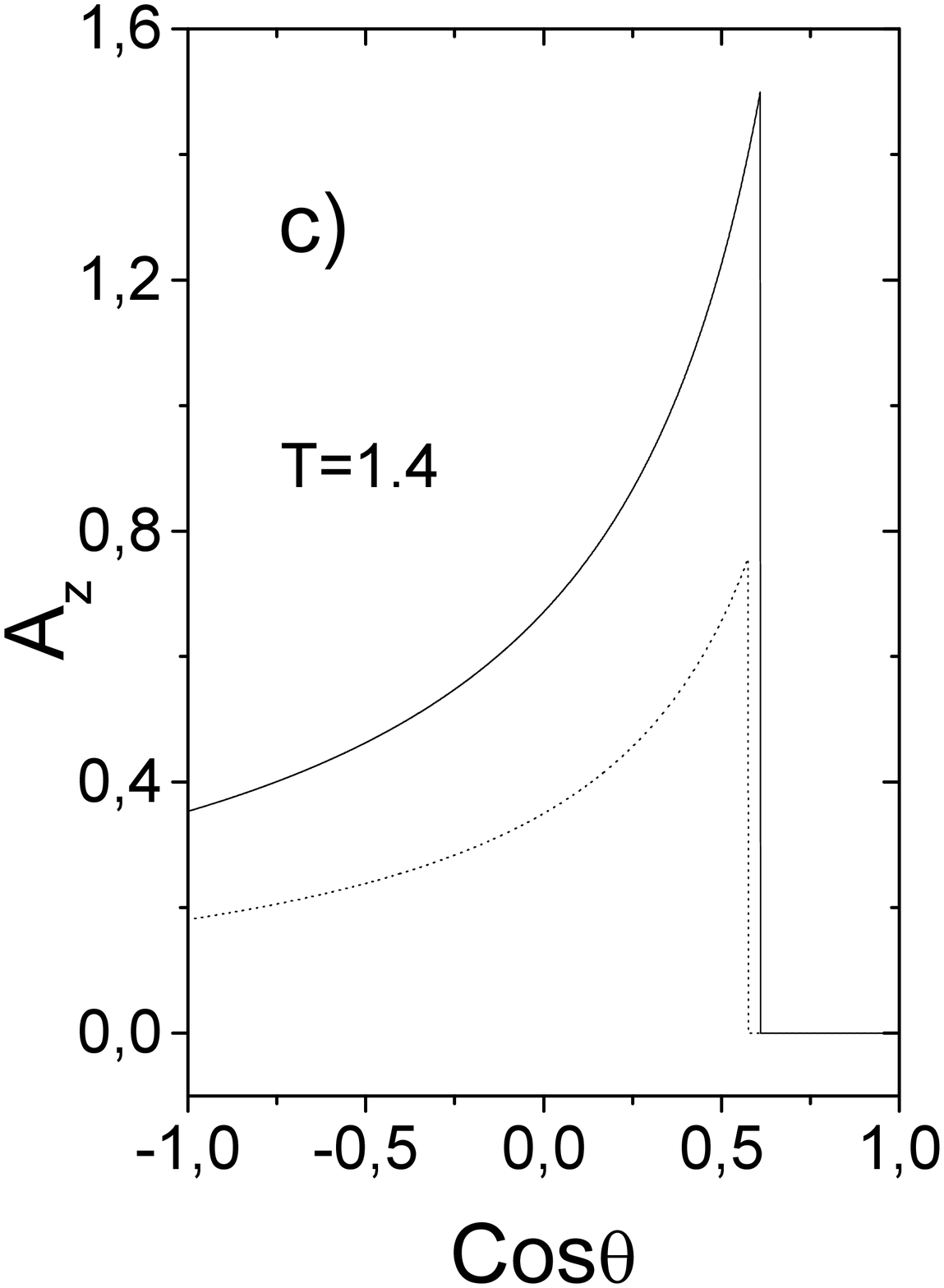}
\caption{Time evolution of BS shock waves for the charge velocity
($\beta=0.7$) less the medium light velocity ($\beta_c=0.75$).
Solid and dashed lines are related to the exact (2.1) and
approximate (2.2) vector potentials.\\
a) BS shock waves fill only back part of S$_{0}$.\\
b) The whole sphere S$_{0}$ is illuminated during some time interval.\\
c) At later times BS again fills only the back part of S$_{0}$.}
\end{figure}
\begin{figure}[h]                       %fig.6
\vspace*{-10mm}
\includegraphics[height=75mm]{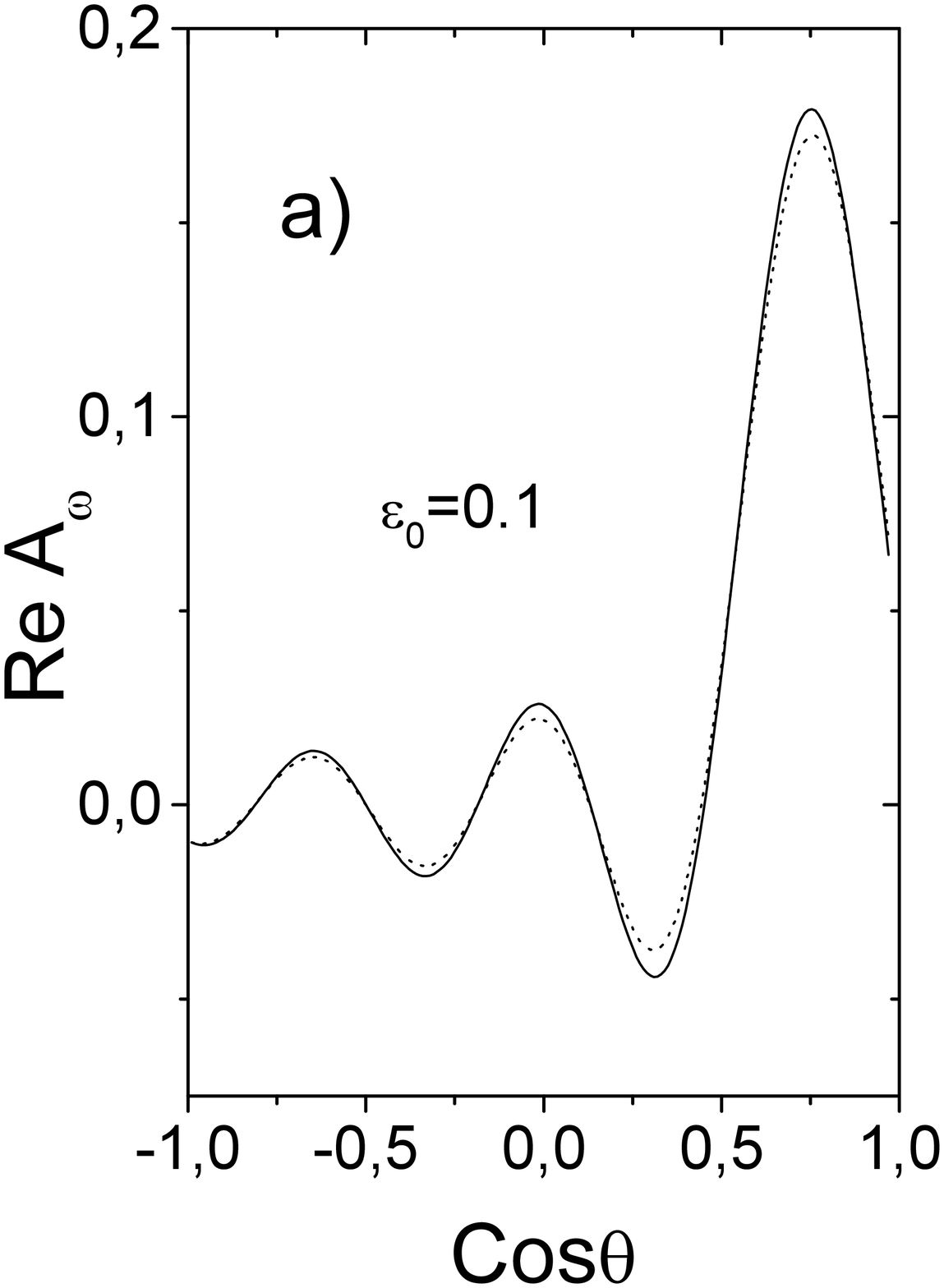}
\includegraphics[height=75mm]{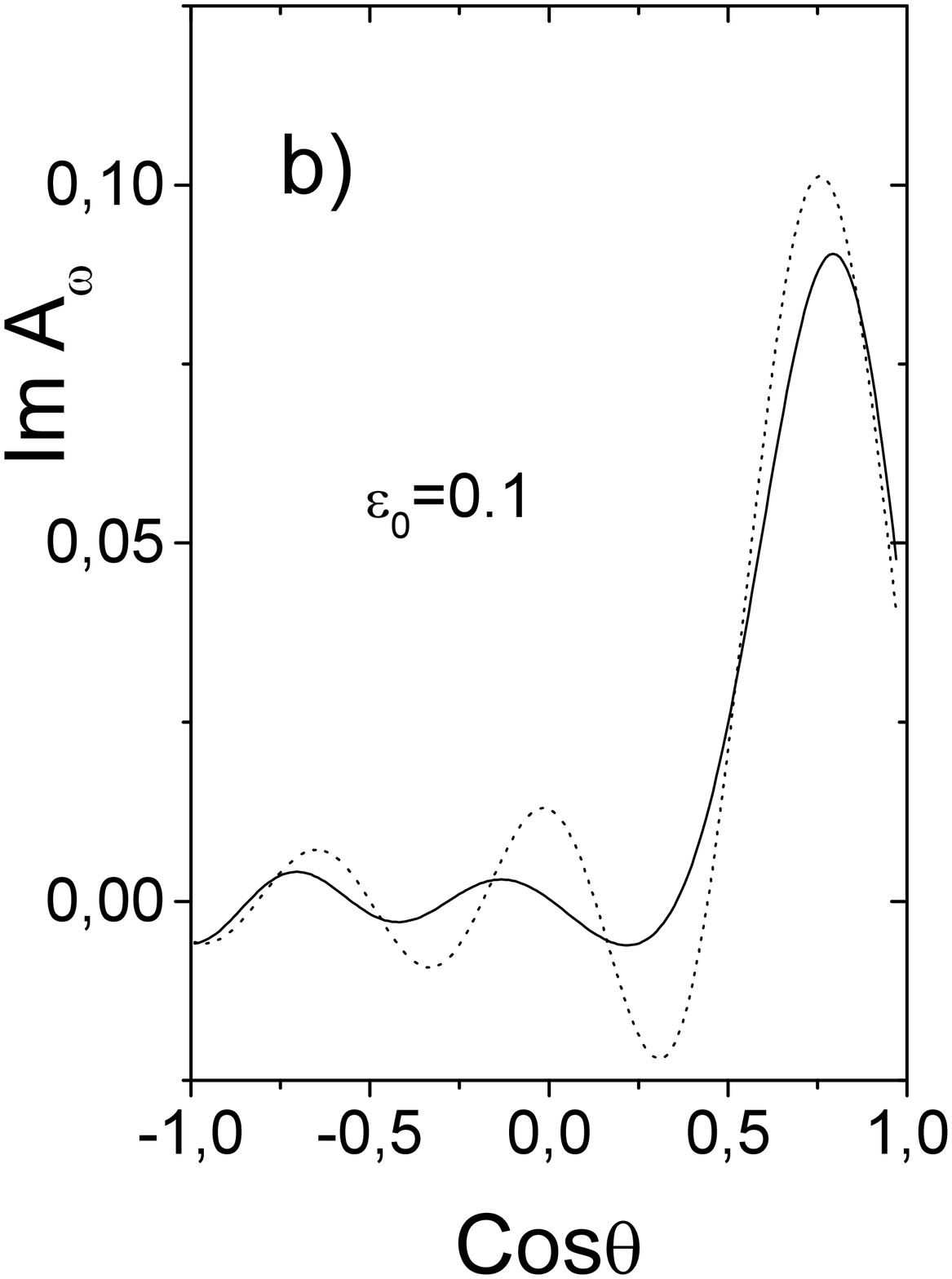}
\includegraphics[height=75mm]{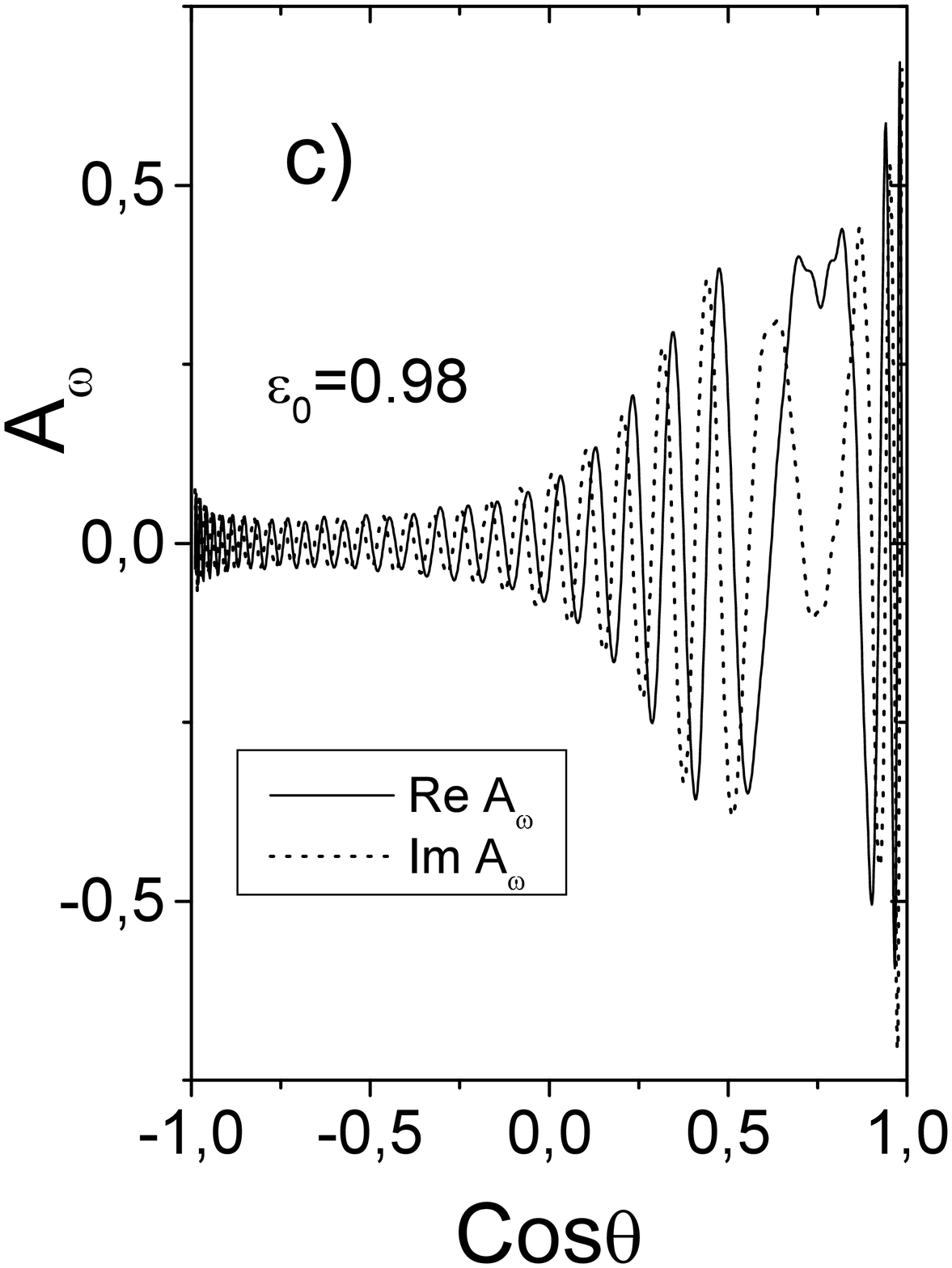}
\caption{The real (a) and  imaginary (b) parts
 of the VP Fourier transform
(in units $e/2\pi c$) on the surface
of S$_0$ for $\epsilon_0=z_0/R_0=0.1$.
The radiation field differs essentially from zero
in the neighborhood of
the \~Cerenkov critical angle $\cos\theta_c=1/\beta_n$.
The solid and dotted curves
refer to the exact and approximate formulae (2.1) and (2.2), resp.
It turns out that a small difference of the Fourier transforms
is responsible for the appearance
of the \~Cerenkov radiation in the space-time representation.\\
c) The real and imaginary parts of $A_\omega$ for
$\epsilon_0=0.98$. The electromagnetic radiation is distributed
over the whole sphere S$_0$.}
\vspace*{-25mm}
\end{figure}

\newpage
\begin{figure}[h]                       %fig.7
%\vspace*{-20mm}
\includegraphics[height=74mm]{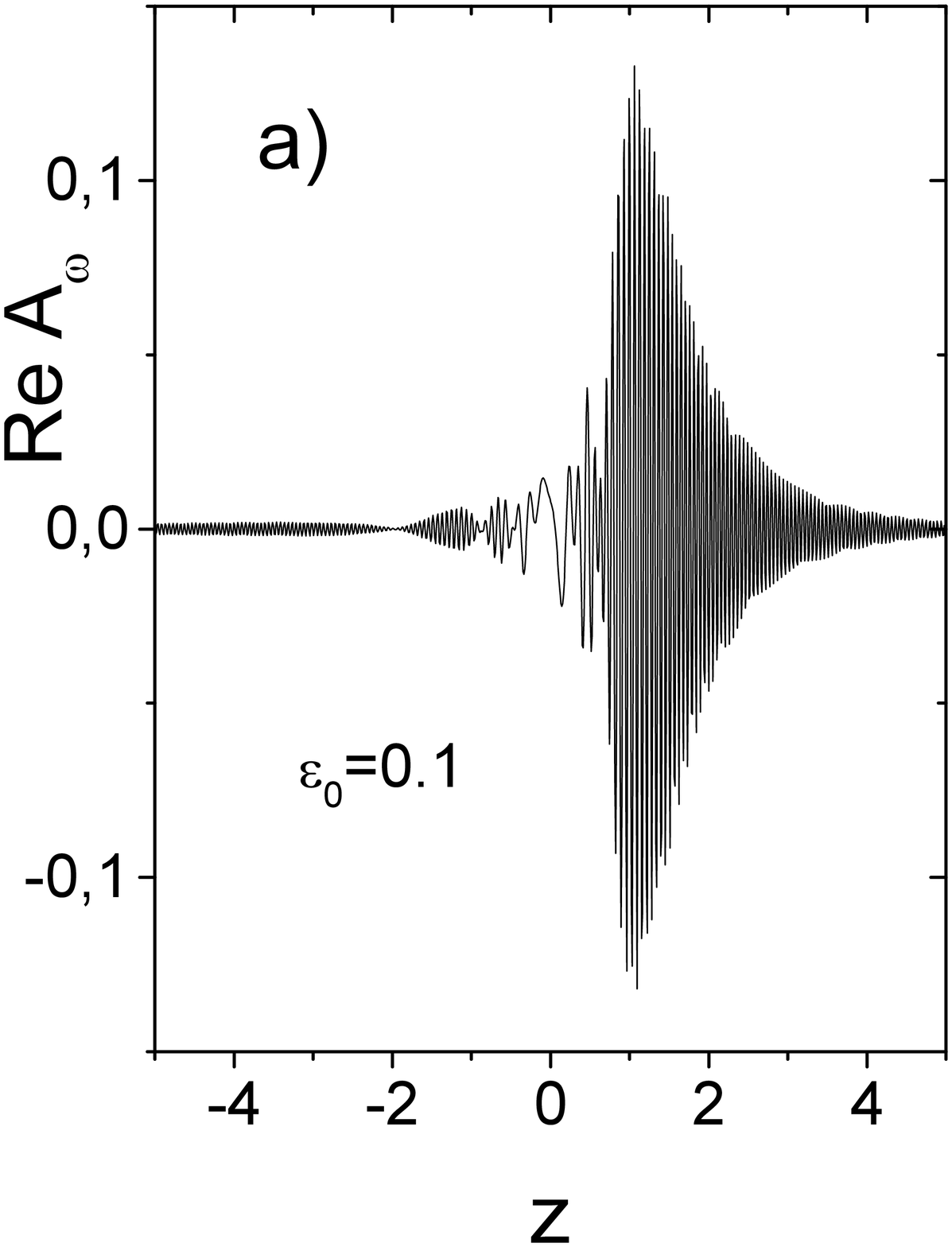}
\includegraphics[height=74mm]{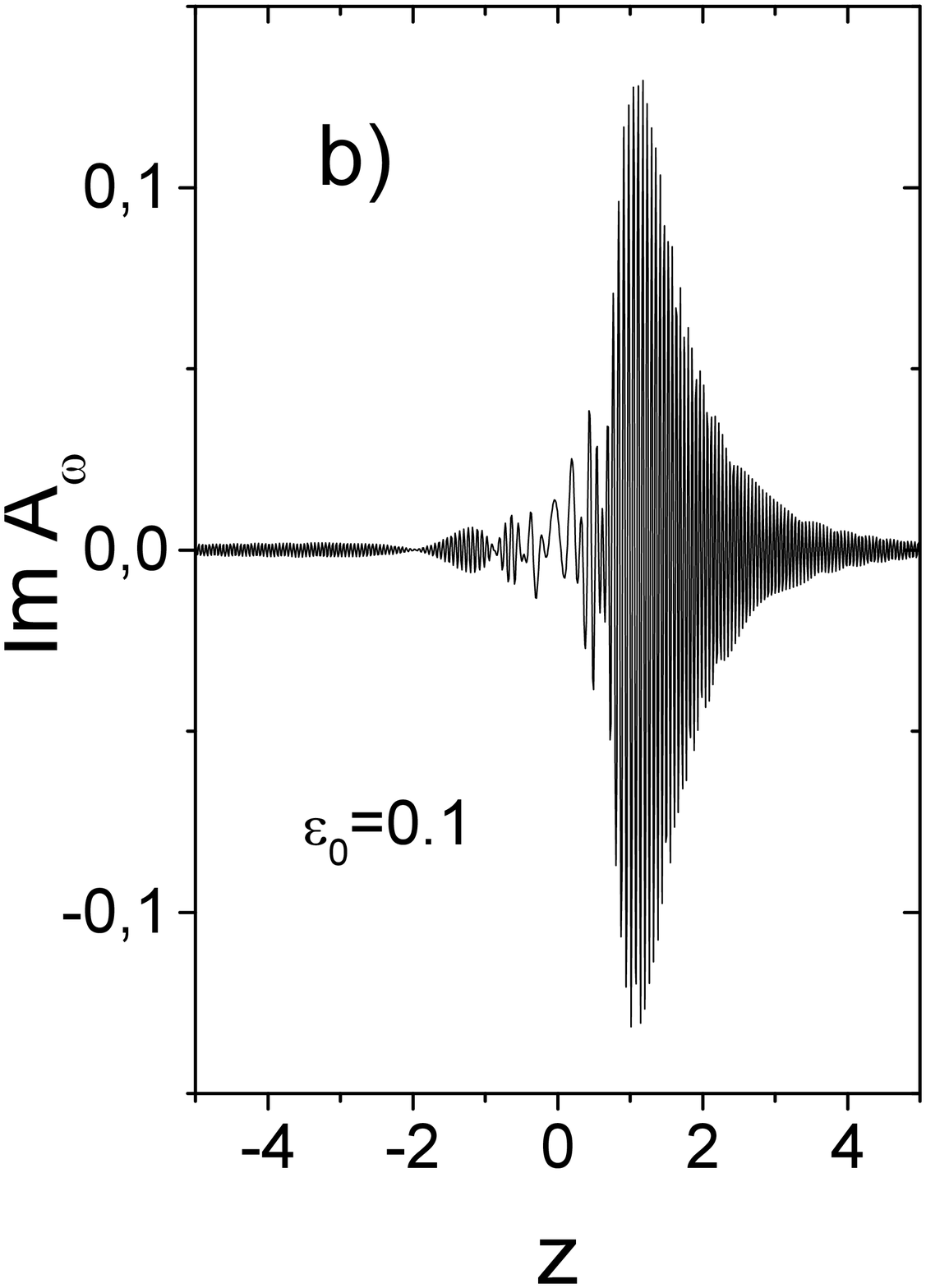}
\includegraphics[height=74mm]{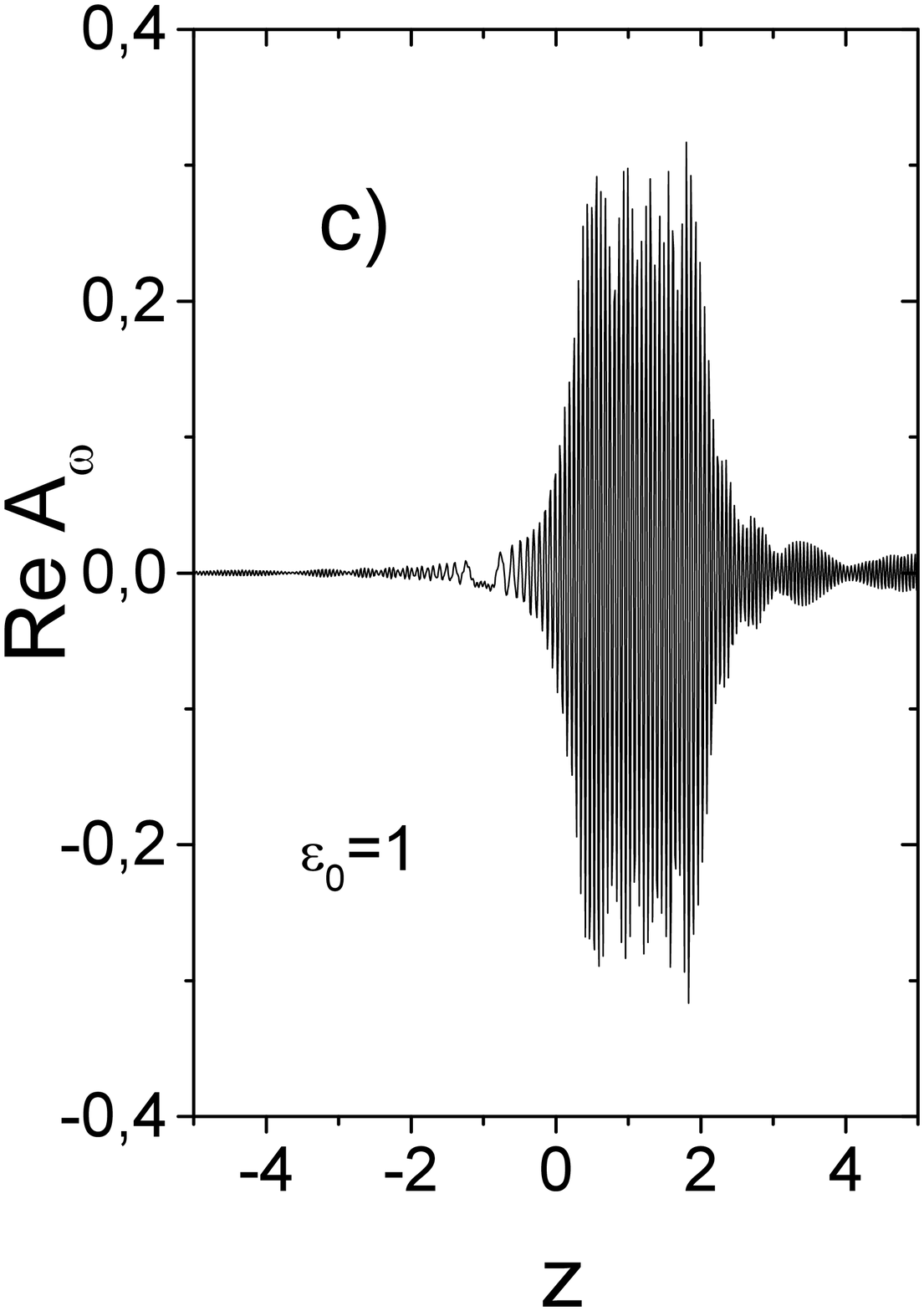}
\caption{The real (a) and imaginary (b) parts of $A_\omega$
on the cylinder C
surface for the ratio of the interval motion to the cylinder radius
$\epsilon_0=0.1$. The electromagnetic radiation differs from zero
in the neighborhood of $z=\gamma_n$,
that corresponds to $\cos\theta_c=1/\beta_n$
on the sphere ($z$ is in units $\rho$, $A_\omega$ in units
$e/2\pi c$).\\
c) The real part $A_\omega$ of  for $\epsilon_0=1$.
There is no sharp radiation
maximum in the neighborhood of $z=\gamma_n$.}
%\vspace*{-25mm}
\end{figure}
%\vspace*{-5mm}
\begin{figure}[h]
%\vspace*{-15mm}
\begin{center}
\includegraphics[height=74mm]{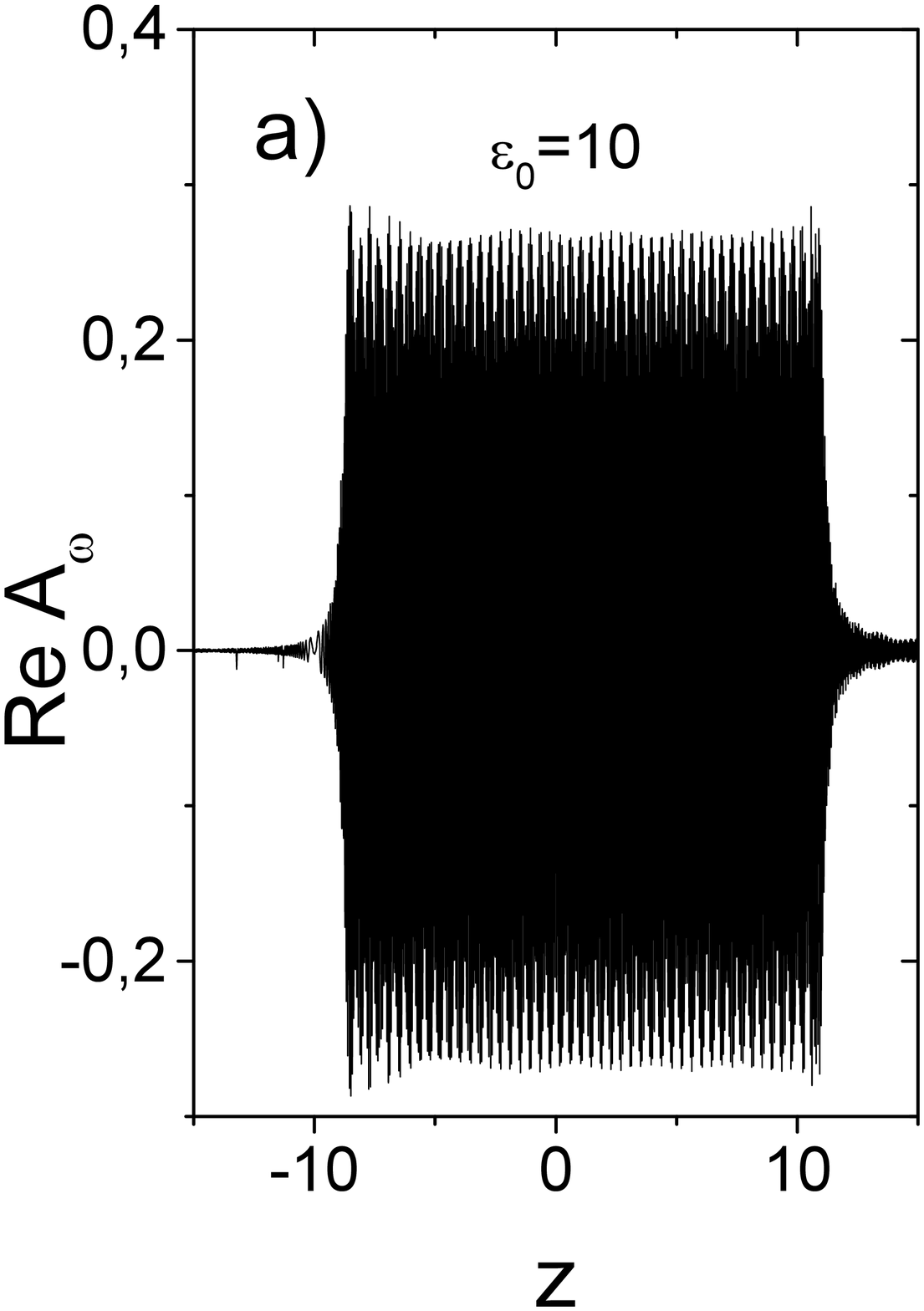}
\includegraphics[height=74mm]{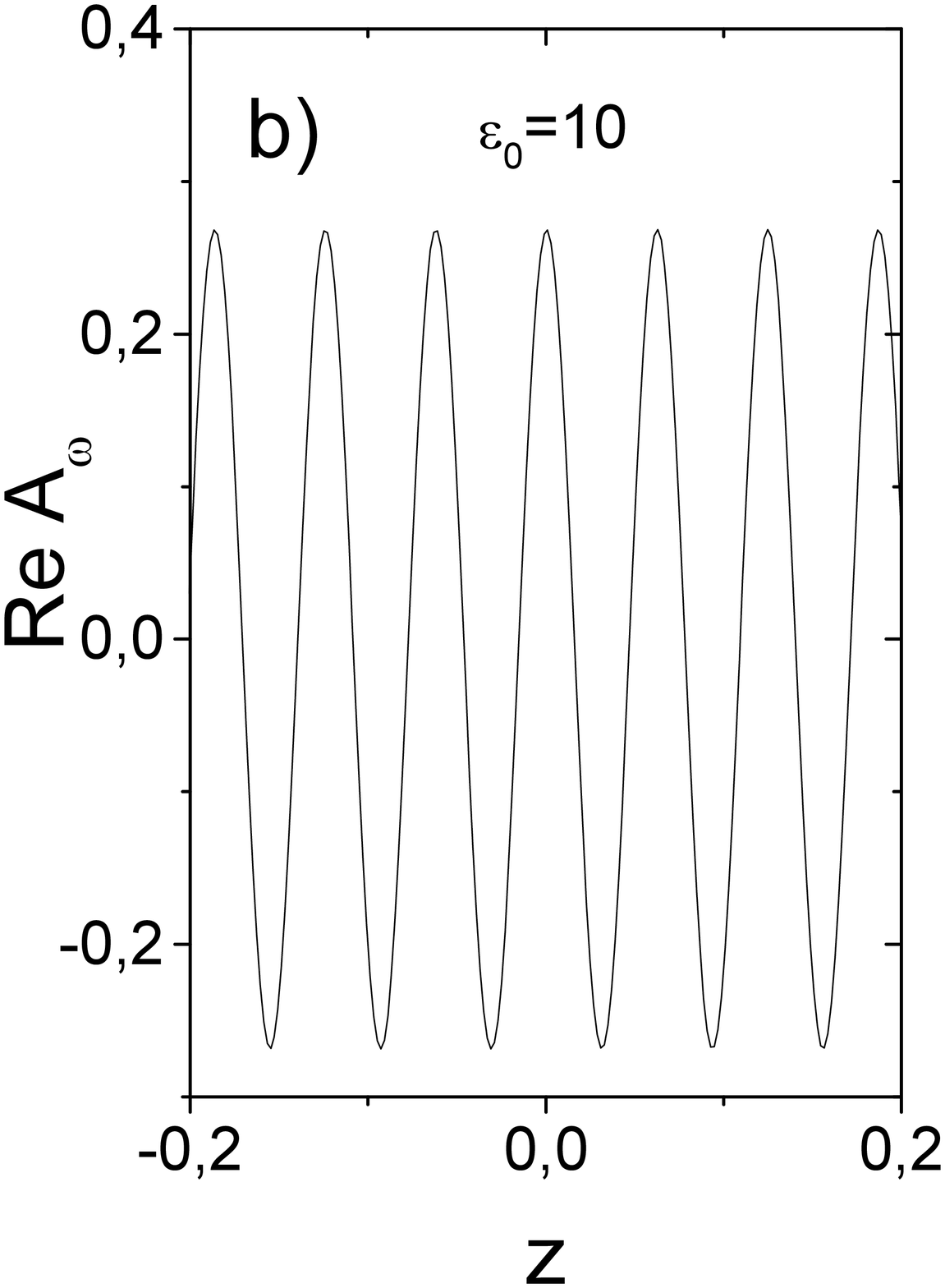}
\end{center}
\caption{The real part of $A_\omega$ for $\epsilon_0=10$.\\
a) There is no radiation
maximum in the neighborhood of $z=\gamma_n$
and the radiation is distributed over the large $z$ interval.\\
b) For the small $z$ interval, $ReA_\omega$
evaluated according to Eq.(4.3)
for $\epsilon_0=10$ and according to Eq.(4.5)
for the infinite motion interval are indistinguishable.}
\vspace*{-45mm}
\end{figure}

\section*{References}
\begin{thebibliography}{25}
\bibitem{1} Heaviside O 1888 {\it Electrician (Nov. 23)} p~83
\item[]\dash 1889 {\it Phil. Mag.} {\bf 27} 324 
\item[]\dash 1912 {\it Electromagnetic Theory} vol.3 
(London: The Electrician) 
\item[]\dash 1971 Repr. ed.: (New York, Chelsea)
\bibitem{2}Cherenkov P A 1934 {\it Dokl. Acad. Nauk SSSR} {\bf 2} 451
\bibitem{3} Tyapkin A A 1974 {\it Usp. Fiz. Nauk} {\bf 112} 731
\bibitem{4} Kaiser  T R 1974 {\it Nature} {\bf 274} 400
\bibitem{5} Frank I M and Tamm I E 1937 {\it Dokl. Acad. Nauk SSSR} 
{\bf 14} 107
\bibitem{6} Tamm I E 1939  {\it J. Phys.  USSR} {\bf 1} No 5-6 439
\bibitem{7} Frank I M 1988 {\it Vavilov-Cherenkov Radiation. 
Theoretical Aspects} (Moscow: Nauka) 
\bibitem{8}  Afanasiev G N and Kartavenko V G 1998
{\it J. Phys. D:  Appl. Phys.} {\bf 31} 2760\\
Afanasiev G N, Kartavenko V G and Magar E N 1999 
{\it Physica B} {\bf 269} 95
\bibitem{9} Landau L D and Lifshitz E M 1992
{\it Electrodynamics of Continuous Media} (Oxford: Pergamon)
\bibitem{10} Lawson J D 1954 {\it Phil. Mag.} {\bf 45} 748
\bibitem{11} Lawson J D 1965  {\it Amer. J. Phys.} {\bf 33} 1002
\bibitem{12} Zrelov V P  and Ruzicka J 1989 {\it Chech. J. Phys. B}
{\bf 39} 368
\bibitem{13} Zrelov V P and Ruzicka J 1992 {\it Chech.J.Phys.} {\bf 42} 45  
\bibitem{14}  Afanasiev G N, Beshtoev Kh and Stepanovsky Yu P 1996
{\it Helv. Phys. Acta} {\bf 69} 111
\bibitem{15} Afanasiev G N, Eliseev S M and Stepanovsky Yu P 1998 
{\it Proc. Roy. Soc. A} {\bf 454} 1049
\bibitem{16} Kobzev A P and Frank I M 1981 {\it Yadern. Fizika} {\bf 334} 134
\bibitem{17} Ginzburg V L and Tsytovich V N 1984 
{\it Transition radiation and transition
scattering} (Moscow: Nauka) (in Russian)
\item[]\dash Ginzburg V L and Tsytovich V N 1979 
{\it Phys. Rep.} {\bf 49} 1
\bibitem{18} Bowler M G 1996 {\it Nucl. Instr. and Methods in Phys. Res. A} 
{\bf 378} 463
\bibitem{19} Ginzburg V L 1940 {\it Zh. Eksp. Teor. Fiz.} {\bf 10} 589 
\item[]\dash 1940 {\it J. Phys. USSR} {\bf 3} 101
\bibitem{20} Akhiezer A I and Berestetzky V B 1981 
{\it Quantum Electrodynamics} ( Moscow: Nauka)
\bibitem{21} Glauber R 1965 {\it in Quantum Optics and Electronics 
(Lectures delivered at Les Houches 1964)} (Eds.: DeWitt C, Blandin A and 
Cohen-Tannoudji C)  (New York:  Gordon and Breach)  pp~93-279
\bibitem{22} Skobeltzyne D V 1975 {\it C.R. Acad. Sci.Paris Ser.B}
{\bf  280} 251 ibid.: 287
\item[]\dash 1977 {\it Usp. Fiz. Nauk} {\bf 122} No~2, 295
\end{thebibliography}
\end{document}